\documentclass[english]{article}
\usepackage[T1]{fontenc}
\usepackage[latin9]{inputenc}
\usepackage{mathrsfs}
\usepackage{amsmath}
\usepackage{amssymb}
\usepackage{esint}

\makeatletter
\usepackage{fullpage}

\makeatother

\usepackage{babel}
\begin{document}
\title{Geometry from geodesics: fine-tuning Ehlers, Pirani, and Schild}
\author{James T. Wheeler\thanks{Utah State University Department of Physics, Logan, UT, USA, jim.wheeler@usu.edu,
orcid 0000-0001-9246-0079}}
\maketitle
\begin{abstract}
Ehlers, Pirani, and Schild argued that measurements of null and timelike
geodesics yield Weyl and projective connections, respectively, with
compatibility in the lightlike limit giving an integrable Weyl connection.
Their conclusions hold only for a 4-dim representation of the conformal
connection on the null cone, and by restricting reparameterizations
of timelike geodesics to yield a torsion-free, affine connection.
An arbitrary connection gives greater freedom. A linear connection
for the conformal symmetry of null geodesics requires the SO(4,2)
representation. The enlarged class of projective transformations of
timelike geodesics changes Weyl's projective curvature, and we find
invariant forms of the torsion and nonmetricity, along with a new,
invariant, second rank tensor field generalizing the dilatational
curvature without requiring a metric. We show that either projective
or conformal connections require a monotonic, twice differentiable
function on a spacetime region foliated by order isomorphic, totally
ordered, twice differentiable timelike curves in a necessarily Lorentzian
geometry. We prove that the conditions for projective and conformal
Ricci flatness imply each other and gauge choices within either can
reduce the geometry to the original Riemannian form. Thus, measurements
of null and timelike geodesics lead to an SO(4,2) connection, with
no requirement for a lightlike limit. Reduction to integrable Weyl
symmetry can only follow from the field equations of a gravity theory.
We show that the simplest quadratic spacetime action leads to this
reduction.
\end{abstract}

\pagebreak{}

\section{Overview}

\subsection{Issues with the Ehlers, Pirani, Schild results}

Attempting to develop the properties of spacetime from observational
data, Ehlers, Pirani, and Schild developed axioms which might allow
us to infer the spacetime connection from geodesic data \cite{EPS original,EPS}.
The basic claim that results is that, by measuring the timelike paths
of a sufficient number of test particles, we may construct a connection
up to the parameterization along the paths. This allows determination
of a projective connection. Similarly, data from lightlike trajectories
determines a conformal connection. Agreement between these connections
in the limit of light speed is shown to lead to a Weyl connection.
Subsequent studies by Matveev and Trautman \cite{Matveev1} and Matveev
and Scholz \cite{Matveev2} showed that the resulting connection must
be integrable Weyl, i.e., the Weyl vector must be a gradient. In this
case there exists a choice of units with vanishing Weyl vector, giving
the appearance\footnote{While the vanishing Weyl vector gauge of Weyl geometry is essentially
Riemannian, Weyl and Riemannian geometries behave differently under
subsequent rescalings.} of a Riemannian geometry.

Our current re-examination, motivated by several issues within these
studies, leads to a different conclusion. The differences arise because
the original and subsequent analyses place a priori constraints on
the form of the connection, changing the conclusions for both the
null and timelike cases.

For the null structure, it was shown by Bateman \cite{Bateman1,Bateman2}
and Cunningham \cite{Cunningham} that the symmetry of electromagnetism
(and therefore, of light cones) is conformal, not Weyl. Because no
real, linear conformal connection exists in 4-dimensions, the restriction
to the form $\Gamma_{\;\;\;\mu\nu}^{\alpha}\;\;\left(\alpha,\mu,\nu\in\left\{ 0,1,2,3\right\} \right)$
automatically restricts to the Weyl symmetry found in \cite{EPS original}.
A proper characterization of the lightlike geodesic symmetries requires
at least the SO$\left(4,2\right)$ connection $\Gamma_{\;\;\;B\nu}^{A}\;\;\left(A,B\in\left\{ 0,\ldots,5\right\} \right)$
of the smallest real, linear representation of the the full conformal
group.

The issues with timelike geodesics are somewhat more involved. The
Ehlers, Pirani, Schild program assumes a torsion-free, affine connection
based on measurement of geodesic curves only, and this implies the
integrable Weyl connection as claimed. However, since the method is
designed to determine the connection from measurements, the assumption
of vanishing torsion seems arbitrary, expecially since generic reparameterizations
change torsion. Although including torsion conflicts with classical
general relativity, it is motivated by Poincarè gauge theory, where
a nonzero spin tensor--as produced by Dirac and gravitino fields
\cite{Wheeler2023a}--provides a source for nonzero torsion, $T_{\;\;\;bc}^{a}-\delta_{b}^{a}T_{\;\;\;ec}^{e}+\delta_{c}^{a}T_{\;\;\;eb}^{e}=\alpha\frac{\delta\mathcal{L}}{\delta\omega_{a}^{\;\;\;bc}}$.

In addition to torsion, a generic connection has additional degrees
of freedom in the nonmetricity tensor. Nonmetricity may be accessed
by arbitrarily defining a symmetric, invertible tensor field $g_{\mu\nu}$
to act as metric (much as introduced in \cite{EPS original}). Then
a general connection will include torsion and nonmetricity. In an
orthonormal 1-form basis $\mathbf{e}^{a}$ with $g_{\mu\nu}=\eta_{ab}e_{\mu}^{\;\;\;a}e_{\nu}^{\;\;\;b}$
we may write
\begin{equation}
\mathbf{d}\mathbf{e}^{a}=\mathbf{e}^{b}\land\boldsymbol{\omega}_{\;\;\;b}^{a}+\mathbf{T}^{a}\label{Torsion structure eq}
\end{equation}
where $\mathbf{T}^{a}$ is the torsion 2-form, and the spin connection
$\boldsymbol{\omega}_{\;\;\;b}^{a}$ may now have a symmetric part,
$\eta_{ac}\boldsymbol{\omega}_{\;\;\;b}^{c}+\eta_{bc}\boldsymbol{\omega}_{\;\;\;a}^{c}$.
Since the metric and connection are \emph{not} required to be compatible,
there is no obstruction to taking $\eta_{ab}$ to match the orthonormal
form of the null light cone metric.

For these and other reasons, the overarching context of our investigations
employs a fully general connection.

Having introduced the field $g_{\mu\nu}$, we then show that \emph{the
conditions required to define a projective connection are identical
to those required to define a conformal connection}. To establish
this result, we look carefully in Section \ref{sec:Reparameterization-in-affine}
at the requirement for, and properties of, a reparameterization function
$\sigma\left(x^{\alpha}\right)$ defined on a suitable region. Further,
we show in Section \ref{sec:Existence-of-} that in dimensions greater
than 2, $\sigma$ must be a monotonic, twice differentiable function
on a region of spacetime foliated by order isomorphic, totally ordered,
$\mathcal{C}^{2}$ timelike curves in a necessarily Lorentzian geometry.
Existence of this function is necessary and sufficient for either
a projective or a conformal connection.

\subsection{Measuring torsion}

The general connection $\boldsymbol{\omega}_{\;\;\;b}^{a}$ cannot
be determined by measurements of geodesics alone, but in the absence
of higher spin sources $\left(\geq3\right)$ it is sufficient to also
measure the torsion \cite{Wheeler 2024b}. With torsion included there
is a 1-parameter family of regional reparameterizations, depending
on the relative weight of the Levi-Civita part of the connection and
the torsion. This generalizes the change under reparameterization
of the asymmetric Thomas connection \cite{Thomas,Trautman} and symmetric
projective connection \cite{EPS original,EPS,Matveev1,Matveev2} studied
previously. Since each member of the enlarged class changes the description
of the geometry in different ways--i.e. the curvature, torsion, nonmetricity,
and metric--and it becomes necessary to generalize the expression
for Weyl's original projectively invariant curvature and to develop
invariant forms of torsion and nonmetricity. We find these forms,
invariant under the full 1-parameter class of projective changes.

In principle, torsion may be measured by studying the curves followed
by spinning and nonspinning particles. Separating a general connection
$\Sigma_{\;\;\;\mu\nu}^{\alpha}$ into symmetric and antisymmetric
pieces, geodesics and parallel transport of the spin vector $s^{\mu}$
satisfy
\begin{eqnarray*}
\frac{du^{\alpha}}{d\tau} & = & -\Sigma_{\;\;\;\left(\mu\nu\right)}^{\alpha}u^{\mu}u^{\nu}\\
\frac{ds^{\alpha}}{d\tau} & = & -\Sigma_{\;\;\;\left(\mu\nu\right)}^{\alpha}s^{\mu}u^{\nu}-\frac{1}{2}T_{\;\;\;\mu\nu}^{\alpha}s^{\mu}u^{\nu}
\end{eqnarray*}
Scalar particles follow geodesics, allowing determination of $\Sigma_{\;\;\;\left(\mu\nu\right)}^{\alpha}$.
Then tracking particles with nonzero spin yields information on the
torsion. The actual situation is more complicated than this, because
even in general relativity spinning particles do not follow geodesics
\cite{Papapetrou I,Papapetrou II}. Furthermore, it is possible to
introduce direct couplings between spin and gravitation, although
experiments do not currently support such terms (see e.g., \cite{Duan et al}).
Without identifying specific measurements, we assume that experiments
can in principle be devised to determine both symmetric and antisymmetric
parts of the connection.

\subsection{A note on conformal symmetry}

We emphasize that our use of the conformal group refers to the preservation
of dimensionless ratios. Unlike some other uses of conformal symmetry
\cite{Barut}, this is an obvious requirement for any type of measurement.
Even in quantum field theory, where quantum corrections may depend
on, say, $\ln p$, this is always relative to some scale, $\ln\frac{p}{p_{0}}$.
Contrary to some previous claims that agreement between the classical
limit of quantum currents and classical geodesics picks a scale, it
is shown in \cite{Wheeler1990}, Section V.C. (and see references
therein) those arguments fail to employ the scale freedom. In fact,
those limits agree while preserving dilatational symmetry of spacetime
as an unmeasurable, arbitrary choice of units even into the quantum
realm. Making such a choice (e.g, by defining the second in terms
of a certain spin flip in Cesium) restores the appearance of a Riemannian
geometry. However, note that the use of local units--such as the
use of red shift as a time coordinate--routinely and correctly occur.

\section{Projective symmetry with torsion}

To establish the freedom to choose either a projective or a conformal
connection, it is necessary to introduce a reparameterization function
$\sigma\left(x^{\alpha}\right)$ on a region of spacetime. We establish
the existence and some of the properties of $\sigma$ in this Section,
and use it to define the 1-parameter class of projective transformations
allowed by the presence of torsion. Writing a connection invariant
under this enlarged class of projections, we find a generalization
of Weyl's original invariant form for the curvature \cite{Weyl 1921},
and an invariant form of the torsion.

In Section 3, we show that we can access the additional degrees of
freedom of a general connection by introducing an arbitrary, nondegenerate,
symmetric tensor field, and provide invariant forms of the tensors
that arise. 

\subsection{Reparameterization versus projection}

Reparameterization of a single curve is the use of a real, monotonic
function $\lambda:\mathbb{R}\rightarrow\mathbb{R}$ to change the
parameter, $\lambda\rightarrow\sigma\left(\lambda\right)$. Requiring
$\sigma\left(\lambda\right)$ to be twice differentiable is then sufficient
to define autoparallels or geodesics. Reparameterization is always
possible, but our interest lies in determining the geometry by measuring
many autoparallels or geodesics. The most we can infer about the geometry
depends on writing the the connection and/or metric so that the resulting
geometry is insensitive to changes of parameter. For this to be possible
we require a \emph{regional reparameterization}--simultaneous reparameterization
of a collection of curves. This is accomplished by using a real function,
$\lambda:\mathbb{R}^{n}\rightarrow\mathbb{R}$ on a neighborhood,
with continuous second derivatives, monotonic on a well-defined class
of curves. Then reparameterization becomes the composition $\sigma\left(\lambda\left(x^{\alpha}\right)\right)$.

\subsection{Projective symmetry in affine geometry\label{sec:Reparameterization-in-affine}}

In an affine geometry $\left(\mathcal{M},\Gamma\right)$ we define
a projective connection as a connection that preserves the autoparallel
equation under regional reparameterization. Consider an arbitrary
connection $\Sigma_{\;\;\;\beta\mu}^{\alpha}$. Combined with a displacement,
$\boldsymbol{\Sigma}_{\;\;\;\beta}^{\alpha}=\Sigma_{\;\;\;\beta\mu}^{\alpha}\mathbf{d}x^{\mu}$,
its 64 ($n^{3}$ in $n$-dimensions) degrees of freedom describe the
general linear transformation $\boldsymbol{\Sigma}_{\;\;\;\beta}^{\alpha}$
of vectors that occurs when comparing vectors in tangent or co-tangent
spaces separated by $\mathbf{d}x^{\mu}$.

First consider reparameterization of a single autoparallel. In a chart,
an autoparallel curve $\mathcal{C}:\mathbb{R}\rightarrow\mathcal{M}$
is given by $x^{\alpha}\left(\lambda\right)$ and has tangent vector
$u^{\alpha}\left(\lambda\right)=\frac{dx^{\alpha}}{d\lambda}$ satisfying
$u^{\alpha}D_{\alpha}u^{\beta}=0$. With the parameterization explicit
this becomes
\begin{equation}
\frac{d^{2}x^{\alpha}}{d\lambda^{2}}=-\Sigma_{\;\;\;\mu\nu}^{\alpha}\frac{dx^{\mu}}{d\lambda}\frac{dx^{\nu}}{d\lambda}\label{Autoparallel}
\end{equation}
Reparameterizing the curve with a function $\sigma\left(\lambda\right):\mathbb{R}\rightarrow\mathbb{R}$,
let $v^{\alpha}=\frac{dx^{\alpha}}{d\sigma}=\frac{d\lambda}{d\sigma}u^{\alpha}$,
and expand $\frac{d}{d\lambda}=\frac{d\sigma}{d\lambda}\frac{d}{d\sigma}$
in Eq.(\ref{Autoparallel}). This introduces a second derivative,
\begin{eqnarray}
\frac{d^{2}x^{\alpha}}{d\sigma^{2}} & = & -\Sigma_{\;\;\;\mu\nu}^{\alpha}\frac{dx^{\mu}}{d\sigma}\frac{dx^{\nu}}{d\sigma}+\frac{d}{d\sigma}\left(\ln\left(\frac{d\lambda}{d\sigma}\right)\right)\frac{dx^{\alpha}}{d\sigma}\label{Reparameterized eq}
\end{eqnarray}
where we use $\frac{1}{\frac{d\sigma}{d\lambda}\frac{d\sigma}{d\lambda}}\frac{d^{2}\sigma}{d\lambda^{2}}=\frac{d}{d\sigma}\left(\ln\left(\frac{d\lambda}{d\sigma}\right)\right)$.
This reparameterization depends only on a monotonic function of a
single variable. The only further requirement for Eq.(\ref{Reparameterized eq})
is for a continuous second derivative, $\frac{d^{2}\sigma}{d\lambda^{2}}$.

Now, following the spirit of the Ehlers, Pirani, Schild program, we
employ a \emph{class} of curves to infer the connection. The change
from reparameterization to a projective connection requires extension
of $\sigma$ to a function on a neighborood. Continuing from (\ref{Reparameterized eq}),
we combine terms on the right side of Eq.(\ref{Reparameterized eq})
to write it in terms of a new connection reproducing the original
form of the autoparallel (\ref{Autoparallel}). The essential difference
between reparameterization of a curve and projective transformation
of the connection lies in using a chart to rewrite the 1-dimensional
derivatives $\frac{d}{d\sigma}\left(\ln\left(\frac{d\sigma}{d\lambda}\right)\right)$
in terms of a spacetime gradient,
\begin{eqnarray*}
\frac{d}{d\sigma}\left(\ln\left(\frac{d\sigma}{d\lambda}\right)\right) & = & \frac{dx^{\mu}}{d\sigma}\frac{\partial}{\partial x^{\mu}}\left(\ln\left(\frac{d\sigma}{d\lambda}\right)\right)=\frac{dx^{\mu}}{d\sigma}\frac{\partial\xi}{\partial x^{\mu}}
\end{eqnarray*}
where we define $\xi\left(x^{\alpha}\right)\equiv\ln\left(\frac{d\sigma\left(x^{\alpha}\right)}{d\lambda}\right)$.
The introduction of the spacetime function $\xi$ allows us to combine
the left side into the altered connection $\tilde{\Sigma}_{\;\;\;\mu\nu}^{\alpha}$
\begin{equation}
\tilde{\Sigma}_{\;\;\;\mu\nu}^{\alpha}\frac{dx^{\mu}}{d\sigma}\frac{dx^{\nu}}{d\sigma}=\left(\Sigma_{\;\;\;\mu\nu}^{\alpha}+\delta_{\mu}^{\alpha}\xi_{,\nu}\right)\frac{dx^{\mu}}{d\sigma}\frac{dx^{\nu}}{d\sigma}\label{Connection after reparameterization}
\end{equation}
so the reparameterized autoparallel takes the same form as the original
in Eq.(\ref{Autoparallel}), but with the new parameter, $\frac{d^{2}x^{\alpha}}{d\sigma^{2}}=-\tilde{\Sigma}_{\;\;\;\mu\nu}^{\alpha}\frac{dx^{\mu}}{d\sigma}\frac{dx^{\nu}}{d\sigma}$.
For Eq.(\ref{Connection after reparameterization}) to hold it is
necessary and sufficient to require
\begin{eqnarray*}
\tilde{\Sigma}_{\;\;\;\left(\mu\nu\right)}^{\alpha} & = & \Sigma_{\;\;\;\left(\mu\nu\right)}^{\alpha}+\frac{1}{2}\left(\delta_{\mu}^{\alpha}\xi_{,\nu}+\delta_{\nu}^{\alpha}\xi_{,\mu}\right)
\end{eqnarray*}
leaving any antisymmetric part of each term undetermined. We parameterize
this remaining indeterminacy by including the most general class of
antisymmetric $\xi_{,\mu}$-dependent forms possible in an affine
geometry. The only available form is a multiple of $\text{\ensuremath{\left(\delta_{\mu}^{\alpha}\xi_{,\nu}-\delta_{\nu}^{\alpha}\xi_{,\mu}\right)}}$,
so the change of a general connection under regional reparameterization
is
\begin{eqnarray}
\tilde{\Sigma}_{\;\;\;\mu\nu}^{\alpha} & = & \Sigma_{\;\;\;\mu\nu}^{\alpha}+a\delta_{\mu}^{\alpha}\xi_{,\nu}+\left(1-a\right)\delta_{\nu}^{\alpha}\xi_{,\mu}\label{Projective transformation}
\end{eqnarray}
for all $a\in\mathbb{R}$. All members of this collection preserve
autoparallels. Equation (\ref{Projective transformation}) includes
both the asymmetric Thomas form \cite{Thomas}, $\tilde{\Sigma}_{\;\;\;\mu\nu}^{\;\alpha}=\Sigma_{\;\;\;\mu\nu}^{\alpha}+\delta_{\mu}^{\alpha}\xi_{,\nu}$
when $a=1$, and the symmetric form $\tilde{\Sigma}_{\;\;\;\mu\nu}^{\;\alpha}=\Sigma_{\;\;\;\mu\nu}^{\alpha}+\frac{1}{2}\left(\delta_{\mu}^{\alpha}\xi_{,\nu}+\delta_{\nu}^{\alpha}\xi_{,\mu}\right)$
used in \cite{EPS original,EPS,Matveev1,Matveev2}when $a=\frac{1}{2}$.

In addition to the autoparallels, the geometry is characterized by
two tensor fields built from the connection--the curvature and the
torsion
\begin{eqnarray}
R_{\;\;\;\beta\mu\nu}^{\alpha} & = & \partial_{\mu}\Sigma_{\;\;\;\beta\nu}^{\alpha}-\partial_{\nu}\Sigma_{\;\;\;\beta\mu}^{\alpha}+\Sigma_{\;\;\;\beta\nu}^{\rho}\Sigma_{\;\;\;\rho\mu}^{\alpha}-\Sigma_{\;\;\;\beta\mu}^{\rho}\Sigma_{\;\;\;\rho\nu}^{\alpha}\label{Curvature}\\
T_{\;\;\;\beta\mu}^{\alpha} & = & \Sigma_{\;\;\;\beta\mu}^{\alpha}-\Sigma_{\;\;\;\mu\beta}^{\alpha}\label{Torsion}
\end{eqnarray}
Substituting Eq.(\ref{Projective transformation}) into the curvature
and torsion, we find dependence on the parameter $a$.

\begin{eqnarray}
\tilde{R}_{\;\;\;\beta\mu\nu}^{\alpha} & = & R_{\;\;\;\beta\mu\nu}^{\alpha}+\left(1-a\right)\left[\xi_{,\beta}T_{\;\;\;\nu\mu}^{\alpha}+\delta_{\nu}^{\alpha}D_{\mu}\xi_{,\beta}-\delta_{\mu}^{\alpha}D_{\nu}\xi_{,\beta}+\left(1-a\right)\left(\delta_{\mu}^{\alpha}\xi_{,\nu}-\delta_{\nu}^{\alpha}\xi_{,\mu}\right)\xi_{,\beta}\right]\label{Projective change in R}\\
\tilde{T}_{\;\;\;\beta\mu}^{\alpha} & = & T_{\;\;\;\beta\mu}^{\alpha}+\left(2a-1\right)\left(\delta_{\beta}^{\alpha}\xi_{,\mu}-\delta_{\mu}^{\alpha}\xi_{,\beta}\right)\label{Projective change in T}
\end{eqnarray}
The dependence of the curvature on the $a=\frac{1}{2}$ symmetric
projection has been shown by Weyl \cite{Weyl 1921} as noted in \cite{Matveev2018},
while Thomas found the antisymmetric $a=1$ expressions. The Thomas
and symmetric cases have opposite effects on the curvature and torsion.
The Thomas form preserves the curvature but not the torsion, while
the symmetric form preserves the torsion but not the curvature. Other
choices for $a$ change both.

While these basic results have been known for a long time, the existence
of the class (\ref{Projective transformation}) of projective changes
shows that the same set of geodesics may be described by distinct
combinations of curvature and torsion. From the Ehlers, Pirani, Schild
point of view, this means that restricting the connection to be independent
of $\xi$ does not yet determine the geometry. We require expressions
for the curvature and torsion which are independent of \emph{both}
$\xi$ and $a$.

Subtracting traces we define the projectively invariant connection
to be
\begin{eqnarray*}
\Pi_{\;\;\;\mu\nu}^{\alpha} & \equiv & \Sigma_{\;\;\;\mu\nu}^{\alpha}-\frac{a}{a\left(n-1\right)+1}\delta_{\mu}^{\alpha}\Sigma_{\;\;\;\beta\nu}^{\beta}-\frac{1-a}{a\left(n-1\right)+1}\delta_{\nu}^{\alpha}\Sigma_{\;\;\;\beta\mu}^{\beta}
\end{eqnarray*}
and verify its invariance, $\tilde{\Pi}_{\;\;\;\mu\nu}^{\;\alpha}=\Pi_{\;\;\;\mu\nu}^{\alpha}$,
under Eq.(\ref{Projective transformation}). Replacing $\Sigma_{\;\;\;\mu\nu}^{\alpha}$
with $\Pi_{\;\;\;\mu\nu}^{\alpha}$ in the definitions (\ref{Curvature})
and (\ref{Torsion}), we immediately find invariant forms of the curvature
and torsion.
\begin{eqnarray}
\hat{\mathcal{R}}_{\;\;\;\beta\mu\nu}^{\alpha} & = & R_{\;\;\;\beta\mu\nu}^{\alpha}+\frac{1-a}{a\left(n-1\right)+1}\left(v_{\beta}T_{\;\;\;\mu\nu}^{\alpha}+\delta_{\mu}^{\alpha}D_{\nu}v_{\beta}-\delta_{\nu}^{\alpha}D_{\mu}v_{\beta}+\frac{1-a}{a\left(n-1\right)+1}\left(\delta_{\mu}^{\alpha}v_{\nu}-\delta_{\nu}^{\alpha}v_{\mu}\right)v_{\beta}\right)\nonumber \\
 &  & +\frac{a}{a\left(n-1\right)+1}\delta_{\beta}^{\alpha}\left(D_{\nu}v_{\mu}-D_{\mu}v_{\nu}+v_{\sigma}T_{\;\;\;\mu\nu}^{\sigma}\right)\label{Projective curvature from Pi}\\
\mathcal{T}_{\;\;\;\mu\nu}^{\alpha} & = & T_{\;\;\;\mu\nu}^{\alpha}+\frac{1-2a}{a\left(n-1\right)+1}\left(\delta_{\mu}^{\alpha}v_{\nu}-\delta_{\nu}^{\alpha}v_{\mu}\right)\label{Projective torsion from Pi}
\end{eqnarray}
where we have set $v_{\nu}\equiv\Sigma_{\;\;\;\beta\nu}^{\beta}$
and $D_{\mu}$ is the initial covariant derivative using $\Sigma_{\;\;\;\beta\nu}^{\alpha}$.

Invariant forms of the torsion and curvature may alternatively be
found by subtracting their traces directly. 
\begin{eqnarray}
\mathcal{R}_{\;\;\;\beta\mu\nu}^{\alpha} & = & R_{\;\;\;\beta\mu\nu}^{\alpha}+\frac{1}{n-1}\left(\delta_{\nu}^{\alpha}R_{\beta\mu}-\delta_{\mu}^{\alpha}R_{\beta\nu}\right)-V_{\beta}\mathcal{T}_{\;\;\;\mu\nu}^{\alpha}\label{Projectively invariant curvature}\\
\mathcal{T}_{\;\;\;\beta\mu}^{\alpha} & = & T_{\;\;\;\beta\mu}^{\alpha}+\frac{1}{n-1}\left(\delta_{\mu}^{\alpha}T_{\;\;\;\nu\beta}^{\nu}-\delta_{\beta}^{\alpha}T_{\;\;\;\nu\mu}^{\nu}\right)\label{Projectively invariant torsion}
\end{eqnarray}
Here $V_{\alpha}=\frac{1-a}{a\left(n-1\right)+1}v_{\alpha}$ transforms
as $\tilde{V}_{\alpha}=V_{\alpha}-\left(1-a\right)\xi_{,\alpha}$.
The curvature expression (\ref{Projectively invariant curvature})
generalizes the vanishing torsion expression $W_{\;\;\;\beta\mu\nu}^{\alpha}=R_{\;\;\;\beta\mu\nu}^{\alpha}+\frac{1}{n-1}\left(\delta_{\nu}^{\alpha}R_{\beta\mu}-\delta_{\mu}^{\alpha}R_{\beta\nu}\right)$
found by Weyl \cite{Weyl 1921}.

There is a discrepancy between the projective change found for the
curvature $\hat{\mathcal{R}}_{\;\;\;\beta\mu\nu}^{\alpha}$ in Eq.(\ref{Projective curvature from Pi})
and $\mathcal{R}_{\;\;\;\beta\mu\nu}^{\alpha}$ in Eq.(\ref{Projectively invariant curvature}).
The second shows the well-known result that curvature is invariant
when $a=1$, while $\hat{\mathcal{R}}_{\;\;\;\beta\mu\nu}^{\alpha}$
still has an additional term
\begin{eqnarray*}
\Phi_{\mu\nu} & \equiv & \frac{a}{a\left(n-1\right)+1}\left(D_{\nu}v_{\mu}-D_{\mu}v_{\nu}+v_{\sigma}T_{\;\;\;\mu\nu}^{\sigma}\right)
\end{eqnarray*}
This reduces to a multiple of $\partial_{\nu}v_{\mu}-\partial_{\mu}v_{\nu}$
which is easily seen to be separately projectively invariant. A detailed
check confirms that
\begin{eqnarray*}
\hat{\mathcal{R}}_{\;\;\;\beta\mu\nu}^{\alpha} & = & \mathcal{R}_{\;\;\;\beta\mu\nu}^{\alpha}-\delta_{\beta}^{\alpha}\Phi_{\mu\nu}
\end{eqnarray*}
Notice that $\Phi_{\mu\nu}$ contributes a nonmetric part to $\hat{\mathcal{R}}_{\;\;\;\beta\mu\nu}^{\alpha}$.

\subsection{Properties of $\sigma\left(x^{\alpha}\right)$ \label{subsec:Properties-of}}

The derivative $\xi_{,\nu}$ appearing in the modification of the
connection requires derivatives of $\sigma\left(\lambda\right)$ away
from the geodesic. Specifically, the gradient of $\xi$ depends on
all partial derivatives of $\frac{d\sigma}{d\lambda}$,
\[
\xi_{,\alpha}=\frac{d\lambda}{d\sigma}\partial_{\alpha}\left(\frac{dx^{\beta}}{d\lambda}\frac{\partial\sigma}{\partial x^{\beta}}\right)
\]
so that continuity of the connection requires continuity of all second
derivatives of $\sigma$. We cannot have a failure of continuity of
the symmetric part of the connection because this would artifically
introduce a Dirac delta singularity in the curvature. We also require
derivatives of the tangent vectors $\frac{dx^{\beta}}{d\lambda}$
in directions away from the curve, another indication that the we
need regional reparameterization.

We may find the order of differentiability of $\sigma\left(x^{\alpha}\right)$
by noting that the geodesic equation determines the torsion-free part
of the connection, and this in turn determines the torsion-free part
of the curvature. Therefore, the order of differentiability depends
on the smoothness of the curvature, which does \emph{not} need to
be continuous. At an abrupt change from matter to free space the energy
tensor drops to zero, giving the Einstein tensor a discontinuity.
It is also possible to have shock waves in gravitating empty spacetime,
leading to a discontinuity in the Weyl curvature. Finally, while the
Bianchi identity $R_{\;\;\;\beta\left[\mu\nu;\sigma\right]}^{\alpha}$
seems to require another derivative, it only depends on first derivatives
of the connection due to the antisymmetrization. We conclude that
there may be discontinuities in the third derivative of $\sigma$. 

Torsion will be discontinuous if there is discontinuity in the spin
tensor arising from a shock wave in the fields that carry spin. However,
since the curvature depends on derivatives of the torsion, this would
lead to curvature singularities. Therefore there is no stronger constraint
on the differentiability of $\sigma$, so we demand continuous partial
second derivatives of $\sigma\left(x^{\alpha}\right)$, and higher
in regions where the curvature is continuous.

The function $\sigma\left(x^{\alpha}\right)$ must also be monotonic
along all timelike geodesics in a specified region. We examine this
condition further in Section \ref{sec:Existence-of-}.

\section{Reparameterization in metric geometry \label{sec:Absorbing-reparameterizations-in}}

A general connection depends on degrees of freedom beyond the Levi-Civita
and contorsion contributions. We may access the remainder of the connection
by choosing an arbitrary, symmetric, nondegenerate tensor as metric.
This extension of the geometry to $\left(\mathcal{M},\Sigma,g\right)$
introduces additional possible forms of the connection, including
conformal. There is no obstruction to introducing an arbitrary metric
as long as we allow nonvanishing nonmetricity\footnote{This choice may be thought of as a gauge freedom of the nonmetricity}.
The metric also allows us to use geodesics as the class of preferred
curves, instead of the autoparallels of affine geometry.

Before studying the effect of regional reparameterizations on metric
geometries we show that, despite working with geodesics, general connections
are possible. In order to keep the metric and connection independent
we generalize from metrics as function-valued mappings $g\left(x\right):TM\times TM\rightarrow\mathbb{R}$,
to consider both equivalence classes of metrics $\left\{ g\cong e^{2\phi}g\right\} $
and metric \emph{functionals}, $g\left[C\right]:TM\times TM\rightarrow\mathbb{R}$.
These are all possible since geodesics are already extrema of the
line element functional. 

Also, we review certain properties of nonmetricity.

\subsection{Geodesics and the connection}

Let the line element be the functional $s\left[C\right]=\int\sqrt{-g_{\alpha\beta}u^{\alpha}u^{\beta}}ds$
where $u^{\alpha}=\frac{dx^{\alpha}}{ds}$ and consider the geometries
that arise when $g$ is taken as a function, an equivalence class
of functions, or a functional.

When the metric is a function of the coordinates $g_{\alpha\beta}\left(x^{\mu}\right)$
and the curve is parameterized by proper time, then varying $\tau=\int\sqrt{-g_{\alpha\beta}u^{\alpha}u^{\beta}}d\tau$
leads uniquely to the Levi-Civita connection. However, using a different
parameter gives the projectively transformed Levi-Civita connection,
as in Eq.(\ref{Connection after reparameterization}). However, the
connection found by varying the line element depends on how the metric
evolves along curves.

When the metric is a member of a conformal equivalence class $\tilde{g}_{\alpha\beta}\cong e^{2\xi}g_{\alpha\beta}\left(x\right)$
an integrable Weyl connection results. For example, Ehlers, Pirani,
and Schild \cite{EPS original} take the metric density

\begin{equation}
\tilde{g}_{\alpha\beta}=\mathfrak{g}_{\alpha\beta}=\frac{g_{\alpha\beta}}{\left(-g\right)^{1/4}}\label{Conformal metric}
\end{equation}
which is easily shown to lead to an integrable Weyl connection. For
a general conformal class $\tilde{g}_{\alpha\beta}=e^{2\xi}g_{\alpha\beta}$,
with $u^{\alpha}=\frac{dx^{\alpha}}{d\tau}$ chosen so that $g_{\alpha\beta}u^{\alpha}u^{\beta}=-1$,
let $\tau=\intop\sqrt{-\tilde{g}_{\alpha\beta}u^{\alpha}u^{\beta}}d\tau$.
Then $\delta\tau=0$ leads to
\begin{eqnarray}
\Sigma_{\;\;\;\mu\nu}^{\alpha} & = & \frac{1}{2}g^{\nu\mu}\left(g_{\mu\beta,\alpha}+g_{\mu\alpha,\beta}-g_{\alpha\beta,\mu}\right)+\left(\delta_{\beta}^{\nu}\xi_{,\alpha}+\delta_{\alpha}^{\nu}\xi_{,\beta}-g^{\nu\mu}g_{\alpha\beta}\xi_{,\mu}\right)\label{Integrable Weyl connection}
\end{eqnarray}
This is the connection of an integrable Weyl geometry with Weyl vector
$W_{\alpha}=\xi_{,\alpha}$. In the case of Eq.(\ref{Conformal metric}),
$\xi_{,\mu}=\partial_{\mu}\ln\sqrt{-g}$.

If the metric is a conformal functional on curves $\tilde{g}_{\alpha\beta}\left[\mathcal{C}\right]=e^{2\int_{C}W_{\mu}dx^{\mu}}g_{\alpha\beta}\left(x\right)$
then the partial derivatives of the metric become $\tilde{g}_{\alpha\beta,\mu}=e^{2\int_{C}W_{\nu}dx^{\nu}}\left(g_{\alpha\beta,\mu}+2g_{\alpha\beta}W_{\mu}\right)$.
This leads to a nonintegrable Weyl geometry whenever $\oint W_{\mu}dx^{\mu}\neq0$.
The connection has the form given in Eq.(\ref{Integrable Weyl connection})
but with the replacement $\xi_{,\alpha}\rightarrow W_{\alpha}$.

Finally, suppose the metric is a general functional on curves, $\tilde{g}_{\alpha\beta}\left[C\right]$.
Then the variation follows the usual pattern but the partial derivatives
of the metric become functional derivatives, $g_{\mu\beta,\alpha}\rightarrow\frac{\delta\tilde{g}_{\mu\beta}}{\delta x^{\alpha}}$.
Setting $\sigma_{\mu\beta\alpha}\equiv\left\langle \frac{\delta\tilde{g}_{\mu\beta}}{\delta x^{\alpha}}\right\rangle $
where the bracket denotes a suitable path average if necessary, we
have a general connection, $\Sigma_{\;\;\;\mu\nu}^{\alpha}=\frac{1}{2}g^{\nu\mu}\left(\sigma_{\mu\beta\alpha}+\sigma_{\mu\alpha\beta}-\sigma_{\alpha\beta\mu}\right)$,
as considered in Section \ref{sec:Reparameterization-in-affine}.

With these possibilities in mind, we consider spacetimes with independent
metric and arbitrary connection $\left(\mathcal{M},\Sigma,g\right)$,
and the independence makes it irrelevant whether the class of special
curves consists of autoparallels or of geodesic curves of a metric
functional. It is then straightforward to restrict the results to
integrable Weyl or Levi-Civita connections if desired.

\subsection{Invariant nonmetricity \label{subsec:Nonmetricity}}

In addition to torsion and curvature, describing the geometry now
involves the metric $g_{\alpha\beta}$, and its covariant derivative,
\begin{eqnarray}
Q_{\alpha\beta\mu} & = & D_{\mu}g_{\alpha\beta}\;\;\;=\;\;\;\partial_{\mu}g_{\alpha\beta}-g_{\rho\beta}\Sigma_{\;\;\;\alpha\mu}^{\rho}-g_{\alpha\rho}\Sigma_{\;\;\;\beta\mu}^{\rho}\label{Nonmetricity}
\end{eqnarray}
the nonmetricity. We need invariant forms for these tensors, since
reparameterization always changes at least one, and generically all
of them. Here we develop projective properties of the nonmetricity.

Equation (\ref{Nonmetricity}) gives the symmetric combination $\Sigma_{\alpha\beta\mu}+\Sigma_{\beta\alpha\mu}$
in terms of nonmetricty and partial derivatives of the metric, while
torsion (\ref{Torsion}) gives the antisymmetric part of the connection.
From these we write a fully general connection by cycling indices
and combining to find
\begin{equation}
\Sigma_{\;\;\;\mu\nu}^{\alpha}=\Gamma_{\;\;\;\mu\nu}^{\alpha}-\frac{1}{2}\left(Q_{\;\;\;\mu\nu}^{\alpha}+Q_{\;\;\;\nu\mu}^{\alpha}-Q_{\mu\nu}^{\;\quad\alpha}\right)+\frac{1}{2}\left(T_{\;\;\;\mu\nu}^{\alpha}+T_{\mu\nu}^{\;\quad\alpha}+T_{\nu\mu}^{\;\quad\alpha}\right)\label{General connection}
\end{equation}
where $\Gamma_{\;\;\;\mu\nu}^{\alpha}$ is the Levi-Civita connection,
satisfying $\partial_{\mu}g_{\alpha\beta}-g_{\rho\beta}\Gamma_{\;\;\;\alpha\mu}^{\rho}-g_{\alpha\rho}\Gamma_{\;\;\;\beta\mu}^{\rho}=0$.

Under the projective class (\ref{Projective transformation}) the
nonmetricity becomes
\begin{equation}
\tilde{Q}_{\alpha\beta\mu}=Q_{\alpha\beta\mu}-2ag_{\alpha\beta}\xi_{\mu}-\left(1-a\right)\left(g_{\beta\mu}\xi_{\alpha}+g_{\alpha\mu}\xi_{\beta}\right)\label{Projective change in Q}
\end{equation}
This change is nonvanishing for all values of $a$. The simplest invariant
form for the nonmetricity is found by subtracting terms formed from
the contraction $W_{\mu}\equiv\frac{1}{2n}Q_{\;\;\;\alpha\mu}^{\alpha}$,
\begin{eqnarray}
\mathcal{Q}_{\;\;\;\beta\mu}^{\alpha} & = & Q_{\;\;\;\beta\mu}^{\alpha}-\frac{n}{a\left(n-1\right)+1}\left(2a\delta_{\beta}^{\alpha}W_{\mu}+\left(1-a\right)\left(\delta_{\mu}^{\alpha}W_{\beta}+g_{\beta\mu}W^{\alpha}\right)\right)\label{Projectively invariant Q}
\end{eqnarray}
By removing both contractions we may write an invariant form that
is also independent of $a$
\begin{eqnarray*}
\mathcal{Q}_{\alpha\beta\mu} & = & Q_{\alpha\beta\mu}+\frac{2n}{\left(n-1\right)\left(n+2\right)}\left(g_{\beta\mu}W_{\alpha}+g_{\mu\alpha}W_{\beta}-\left(n+1\right)g_{\alpha\beta}W_{\mu}\right)\\
 &  & -\frac{2n}{\left(n-1\right)\left(n+2\right)}\left(g_{\beta\mu}nU_{\alpha}+g_{\mu\alpha}nU_{\beta}-2g_{\alpha\beta}U_{\mu}\right)
\end{eqnarray*}
where we define the second independent contraction of the nonmetricity
as $U_{\mu}=\frac{1}{2n}Q_{\;\;\;\mu\alpha}^{\alpha}$.

The contraction $W_{\mu}=\frac{1}{2n}Q_{\;\;\;\alpha\mu}^{\alpha}$
is the Weyl vector. To see this note that if the nonmetricity is pure
trace $Q_{\;\;\;\beta\mu}^{\alpha}=\frac{1}{n}\delta_{\beta}^{\alpha}Q_{\;\;\;\nu\mu}^{\nu}$,
then the metric satisfies the defining equation $D_{\mu}g_{\alpha\beta}=2g_{\alpha\beta}W_{\mu}$
for Weyl geometry. Substituting $Q_{\;\;\;\beta\mu}^{\alpha}=\frac{1}{n}\delta_{\beta}^{\alpha}Q_{\;\;\;\nu\mu}^{\nu}=2\delta_{\beta}^{\alpha}W_{\mu}$
into Eq.\ref{General connection} yields a Weyl connection with torsion.
\begin{eqnarray*}
\Sigma_{\;\;\;\mu\nu}^{\alpha} & = & \Gamma_{\;\;\;\mu\nu}^{\alpha}-\left(\delta_{\mu}^{\alpha}W_{\nu}+\delta_{\nu}^{\alpha}W_{\mu}-g_{\mu\nu}W^{\alpha}\right)+\frac{1}{2}\left(T_{\;\;\;\mu\nu}^{\alpha}+T_{\mu\nu}^{\;\quad\alpha}+T_{\nu\mu}^{\;\quad\alpha}\right)
\end{eqnarray*}
As a gauge vector of dilatations, the Weyl vector changes as
\begin{equation}
\tilde{W}_{\alpha}=W_{\alpha}+\partial_{\alpha}\phi\label{Conformal dependence of W}
\end{equation}
leaving covariant derivatives invariant under \emph{conformal} rescalings.
However, here we are considering \emph{projective} transformations.

Under \emph{projective} transformation, Eq.(\ref{Projective change in Q}),
shows that the trace of the nonmetricity satisfies $\tilde{Q}_{\;\;\;\alpha\mu}^{\alpha}=Q_{\;\;\;\alpha\mu}^{\alpha}-2\left(\left(n-1\right)a+1\right)\xi_{,\mu}$
so we have
\begin{eqnarray}
W_{\mu} & \equiv & \frac{1}{2n}Q_{\;\;\;\alpha\mu}^{\alpha}\label{Weyl vector from Q}\\
\tilde{W}_{\mu} & = & W_{\mu}-\frac{1}{n}\left(a\left(n-1\right)+1\right)\xi_{,\mu}\label{Gauge dependence of W}
\end{eqnarray}
An integrable Weyl vector, $W_{\mu}=\partial_{\mu}\zeta$, may therefore
be removed by \emph{either} projective or conformal transformation.

The projective change in $U_{\mu}$ is slightly different, $\tilde{U}_{\beta}=U_{\beta}+\frac{1}{2n}\left(a\left(n-1\right)-\left(n+1\right)\right)\xi_{\beta}$.

In the next two Sections we explore the projectively (\ref{subsec:Projectively-invariant-metric})
and conformally (\ref{subsec:Conformal-invariance}) invariant geometries
within this extended $\left(M,\Gamma,g\right)$ context.

\section{Two geometrizations: projective and conformal\label{sec:Effects-of-projective}}

A projective transformation of the connection absorbs the change in
tangent vectors in such a way that the geodesic equation has the same
form in any parameterization. However, when there is a metric we find
another, equally valid, way to accomplish the same goal. A distinct
approach arises because the projective transformation (\ref{Projective transformation})
also affects inner products.

To see this, first consider how independent reparameterizations affect
inner products of tangent vectors. Let $x^{\alpha}\left(\lambda_{1}\right)$
and $x^{\alpha}\left(\lambda_{2}\right)$ describe two curves with
tangents $u^{\alpha}=\frac{dx^{\alpha}}{d\lambda_{1}}$ and $v^{\alpha}=\frac{dx^{\alpha}}{d\lambda_{2}}$.
We may compute the inner product of these vectors at any intersection
point of the curves, $\left[g_{\alpha\beta}u^{\alpha}v^{\beta}\right]\left(\mathcal{P}\right)$.
Reparameterizing the curves by $\sigma_{1}\left(\lambda_{1}\right)$
and $\sigma_{2}\left(\lambda_{2}\right)$ respectively, the inner
product changes to
\begin{equation}
g_{\alpha\beta}u^{\alpha}v^{\beta}=\left(\frac{d\sigma_{1}}{d\lambda_{1}}\right)\left(\frac{d\sigma_{2}}{d\lambda_{2}}\right)g_{\alpha\beta}\frac{dx^{\alpha}}{d\sigma_{1}}\frac{dx^{\beta}}{d\sigma_{2}}\label{Reparametrization change in guv}
\end{equation}
Next, let $\sigma\left(\lambda\right)$ become a function on a neighborhood
as in Section \ref{sec:Reparameterization-in-affine}. We may expand
any curve $C\left(\lambda\right)$ in a neighborhood as
\begin{eqnarray*}
C\left(x\right)=x^{\alpha}\left(\lambda\left(x\right)\right) & = & x_{0}^{\alpha}+\frac{dx_{1}^{\alpha}\left(\lambda\right)}{d\lambda}\left(\mathcal{P}\right)d\lambda\\
 & = & x_{0}^{\alpha}+u^{\alpha}\left(\mathcal{P}\right)\left(\lambda-\lambda\left(\mathcal{P}\right)\right)
\end{eqnarray*}
Then, changing the parameter we have
\begin{eqnarray*}
\frac{dx^{\alpha}}{d\sigma}\left(\mathcal{P}\right) & = & \left[\frac{d}{d\sigma}\left(x_{0}^{\alpha}+u^{\alpha}\left(\mathcal{P}\right)\left(\lambda-\lambda\left(\mathcal{P}\right)\right)\right)\right]_{\mathcal{P}}\\
 & = & u^{\alpha}\left.\frac{d\lambda}{d\sigma}\left(x\right)\right|_{\mathcal{P}}\\
 & = & e^{\xi}\left(\mathcal{P}\right)v^{\alpha}
\end{eqnarray*}
so that the factors in Eq.(\ref{Reparametrization change in guv})
become identical, giving 
\[
g_{\alpha\beta}\frac{dx^{\alpha}}{d\lambda}\frac{dx^{\beta}}{d\lambda}=e^{2\xi}\left(g_{\alpha\beta}\frac{dx^{\alpha}}{d\sigma}\frac{dx^{\beta}}{d\sigma}\right)
\]

There are now two routes to absorb projections into the geometry,
depending on how we treat the inner product. Writing $\tilde{u}^{\alpha}=e^{\xi}u^{\alpha}$
and holding the metric fixed, we find the connection altered as before.
However, we also have the option of associating the factor $e^{2\xi}$
with the metric,
\begin{equation}
g_{\alpha\beta}\tilde{v}^{\alpha}\tilde{v}^{\beta}=\tilde{g}_{\alpha\beta}v^{\alpha}v^{\beta}\label{Transformed metric}
\end{equation}
In this case

\begin{equation}
\tilde{g}_{\alpha\beta}=e^{2\xi}g_{\alpha\beta}\label{Conformal transformation of the metric}
\end{equation}
gives a \emph{conformal} equivalence class of metrics. As a result,
we now have two closely related but distinct ways describe the geometry:
fixed metric with projective connection, or conformal metric with
invariant connection. We consider each case, tracking the consequent
changes in torsion, nonmetricity, and curvature.

In Section \ref{sec:Existence-of-} we consider topological requirements
on parameters that allow this change to be absorbed into either the
metric or the connection, and show that\emph{ either} geometrization
of projections makes the \emph{same} implicit assumption about the
existence and properties of $\sigma\left(x^{\alpha}\right)$.

\subsection{Projectively invariant metric geometry \label{subsec:Projectively-invariant-metric}}

With fixed metric, the forms of the curvature and torsion invariant
under (\ref{Projective transformation}) remain unchanged from Eqs.(\ref{Projective change in R})
and (\ref{Projective change in T}) above. In addition we have found
the new invariant $\Phi_{\alpha\beta}$ and an invariant form of the
nonmetricity. The invariant expressions depend on vectors $V_{\alpha}$
and $W_{\alpha}$, but since the projective changes in these are proportional,
\begin{eqnarray*}
\tilde{V}_{\alpha} & = & V_{\alpha}-\left(1-a\right)\xi_{,\alpha}\\
\tilde{W}_{\alpha} & = & W_{\alpha}-\frac{1}{n}\left(a\left(n-1\right)+1\right)\xi_{,\alpha}
\end{eqnarray*}
we may write all four equations in terms of a single vector. Defining
\begin{eqnarray*}
\mathcal{W}_{\alpha} & \equiv & \frac{n}{a\left(n-1\right)+1}W_{\alpha}
\end{eqnarray*}
we have $\tilde{\mathcal{W}}_{\alpha}\equiv\mathcal{W}_{\alpha}-\xi_{,\alpha}$
and we may replace $V_{\alpha}$ by $\left(1-a\right)\mathcal{W}_{\alpha}$.
Then the geometry invariant under projective changes in $\left(a,\xi\right)$
may be written as
\begin{eqnarray}
\mathcal{R}_{\;\;\;\beta\mu\nu}^{\alpha} & = & R_{\;\;\;\beta\mu\nu}^{\alpha}+\frac{1}{n-1}\left(\delta_{\nu}^{\alpha}R_{\beta\mu}-\delta_{\mu}^{\alpha}R_{\beta\nu}\right)-\left(1-a\right)\mathcal{W}_{\alpha}\mathcal{T}_{\;\;\;\mu\nu}^{\alpha}\nonumber \\
\mathcal{T}_{\;\;\;\beta\mu}^{\alpha} & = & T_{\;\;\;\beta\mu}^{\alpha}+\frac{1}{n-1}\left(\delta_{\mu}^{\alpha}T_{\;\;\;\nu\beta}^{\nu}-\delta_{\beta}^{\alpha}T_{\;\;\;\nu\mu}^{\nu}\right)\nonumber \\
\Phi_{\mu\nu} & = & D_{\nu}\mathcal{W}_{\mu}-D_{\mu}\mathcal{W}_{\nu}+\mathcal{W}_{\sigma}T_{\;\;\;\mu\nu}^{\sigma}\nonumber \\
\mathcal{Q}_{\;\;\;\beta\mu}^{\alpha} & = & Q_{\;\;\;\beta\mu}^{\alpha}-\left(2a\delta_{\beta}^{\alpha}\mathcal{W}_{\mu}+\left(1-a\right)\left(\delta_{\mu}^{\alpha}\mathcal{W}_{\beta}+g_{\beta\mu}\mathcal{W}^{\alpha}\right)\right)\label{Tensors of projective geometry}
\end{eqnarray}

We now consider consequences of the conformal equivalence class of
metrics, which also depend on $W_{\alpha}$, but now with conformal
gauging given by Eq.(\ref{Conformal dependence of W}).

\subsection{Conformal invariant geometry \label{subsec:Conformal-invariance}}

In this Subsection we allow the metric to vary conformally as in Eq.(\ref{Transformed metric})
with reparameterizations expressed in terms of the equivalence class
$\left\{ \tilde{g}_{\alpha\beta}\cong g_{\alpha\beta}\left|\tilde{g}_{\alpha\beta}=e^{2\xi}g_{\alpha\beta}\right.\right\} $.
These conformal transformations depend on the \emph{same} $C^{2}$
functions $\xi$ defined for projective transformations, and give
the same reparameterizations of curves. However invariance is now
accomplished within integrable Weyl geometry. Specific requirements
on $\xi$ are discussed in Section \ref{sec:Existence-of-}. Also
observe that the dilatations of Weyl geometry do not exhaust the conformal
transformations, which are defined by the equivalence relation \ref{Conformal transformation of the metric}.
This issue is addressed in detail in Sections \ref{sec:Linear-representation-of}
and \ref{sec:Spacetime-with-a}.

Within the conformal equivalence class $\left\{ \tilde{g}_{\alpha\beta}\left|\tilde{g}_{\alpha\beta}=e^{2\xi}g_{\alpha\beta}\right.\right\} $
there must exist one element satisfying $g_{\alpha\beta}u^{\alpha}u^{\beta}=-1$,
where the tangent vector $u^{\alpha}$ has parameter $\tau$. For
this $g_{\alpha\beta}$ the geodesic equation has the autoparallel
form,
\begin{eqnarray*}
\frac{du^{\nu}}{d\tau} & = & -\Gamma_{\;\;\;\alpha\beta}^{\nu}u^{\alpha}u^{\beta}
\end{eqnarray*}
where for the moment we let $g_{\alpha\beta}=g_{\alpha\beta}\left(x^{\mu}\right)$
so we have the Levi-Civita form $\Gamma_{\;\;\;\alpha\beta}^{\nu}=\frac{1}{2}g^{\nu\mu}\left(g_{\mu\alpha,\beta}+g_{\mu\beta,\alpha}-g_{\alpha\beta,\mu}\right)$.
Substituting any other member of the class, $\tilde{g}_{\alpha\beta}=e^{2\xi}g_{\alpha\beta}$
we find
\begin{eqnarray*}
\frac{du^{\nu}}{d\tau} & = & -\left(\Gamma_{\;\;\;\alpha\beta}^{\nu}+\left(\delta_{\alpha}^{\nu}\xi_{,\beta}+\delta_{\beta}^{\nu}\xi_{,\alpha}-g_{\alpha\beta}\xi^{,\nu}\right)\right)u^{\alpha}u^{\beta}
\end{eqnarray*}
so we define the equivalence class of connections $\tilde{\Gamma}_{\;\;\;\alpha\beta}^{\nu}\cong\hat{\Gamma}_{\;\;\;\alpha\beta}^{\nu}$
iff there exists $\zeta$ such that
\begin{eqnarray*}
\tilde{\Gamma}_{\;\;\;\alpha\beta}^{\nu} & = & \hat{\Gamma}_{\;\;\;\alpha\beta}^{\nu}+\left(\delta_{\alpha}^{\nu}\zeta_{,\beta}+\delta_{\beta}^{\nu}\zeta_{,\alpha}-g_{\alpha\beta}\zeta^{,\nu}\right)
\end{eqnarray*}
For a general member of the class, the nonmetricity is given by the
pure-trace form
\begin{eqnarray*}
\tilde{Q}_{\alpha\beta\mu} & = & g_{\alpha\beta,\mu}-\tilde{\Gamma}_{\alpha\beta\mu}-\tilde{\Gamma}_{\beta\alpha\mu}=-2g_{\alpha\beta}\xi_{,\mu}
\end{eqnarray*}
and we may write
\begin{eqnarray*}
Q_{\;\;\;\alpha\mu}^{\alpha} & = & -2n\xi_{,\mu}
\end{eqnarray*}
which shows that the Weyl vector $\frac{1}{2n}Q_{\;\;\;\alpha\mu}^{\alpha}$
is integrable. The connection, torsion, nonmetricity, and curvature
all change with the conformal change of the metric. These changes
are offset by forming the conformally invariant connection.

\subsubsection{Conformally invariant connection}

Weyl geometry, which adds dilatational symmetry to the Poincarè symmetry
of spacetime, has been studied for over a century (see \cite{Wheeler2018a}
and references therein). Here we briefly relate the Weyl structure
to the general connections we have been examining. 

In accordance with the discussion in Subsection (\ref{subsec:Nonmetricity})
we need only subtract out the trace of the nonmetricity to make it
conformally invariant. Together with the Levi-Civita part of an arbitrary
connection we see that the metric derivatives combine with the corresponding
Weyl vector subtractions to give the dilatationally invariant derivative
\begin{equation}
\mathscr{D}_{\mu}g_{\nu\beta}=g_{\nu\beta,\mu}-2g_{\nu\beta}W_{\mu}\label{Dilatationally invariant derivative}
\end{equation}
where the derivative of the conformal factor $e^{2\xi}$ is cancelled
by the \emph{conformal} change in the Weyl vector, $\tilde{W}_{\mu}=W_{\mu}+\xi_{,\mu}$.
The combination of the Levi-Civita and Weyl terms in the covariant
derivative now combine to give
\[
\Psi_{\;\;\;\mu\nu}^{\alpha}=\frac{1}{2}g^{\alpha\beta}\left(\mathscr{D}_{\nu}g_{\mu\beta}+\mathscr{D}_{\mu}g_{\nu\beta}-\mathscr{D}_{\beta}g_{\mu\nu}\right)
\]
and the general invariant connection becomes
\begin{eqnarray}
\Phi_{\;\;\;\mu\nu}^{\alpha} & = & \Psi_{\;\;\;\mu\nu}^{\alpha}-\frac{1}{2}\left(\hat{Q}_{\;\;\;\mu\nu}^{\alpha}+\hat{Q}_{\;\;\;\nu\mu}^{\alpha}-\hat{Q}_{\mu\nu}^{\;\quad\alpha}\right)+\frac{1}{2}\left(T_{\;\;\;\mu\nu}^{\alpha}+T_{\mu\nu}^{\;\quad\alpha}+T_{\nu\mu}^{\;\quad\alpha}\right)\label{Weyl connection}
\end{eqnarray}
where $\hat{Q}_{\;\;\;\mu\nu}^{\alpha}$ is any remaining traceless
part of the nonmetricity. The full connection $\Phi_{\;\;\;\mu\nu}^{\alpha}$
is now invariant under conformal transformation, so the curvature,
torsion, and residual nonmetricity are also conformally invariant,
although the curvature now includes terms dependent on the Weyl vector.
An integrable Weyl vector, removable by a conformal transformation,
is sufficient to accommodate any rescaling $e^{2\xi}$. By taking
$\xi\left(x^{\alpha}\right)\equiv\ln\left(\frac{d\sigma\left(x^{\alpha}\right)}{d\lambda}\right)$,
curves are reparameterized in the same way as for the corresponding
projective transformation.

\subsubsection{Invariant torsion, nonmetricity, and curvature}

The Weyl connection is dilatationally invariant, so the curvature,
torsion, and traceless nonmetricity that follow from it are all invariant
as well. To see the dependence of each field on the Weyl vector, rewrite
Eq.(\ref{Weyl connection}) as
\begin{eqnarray*}
\Phi_{\;\;\;\beta\mu}^{\alpha} & = & \Gamma_{\;\;\;\beta\mu}^{\alpha}-\left(\delta_{\beta}^{\alpha}W_{\mu}+\delta_{\mu}^{\alpha}W_{\beta}-g_{\beta\mu}W^{\alpha}\right)+\Delta_{\;\;\;\beta\mu}^{\alpha}
\end{eqnarray*}
where the conformally invariant $\Delta_{\;\;\;\beta\mu}^{\alpha}$
includes the contorsion of $T_{\;\;\;\mu\nu}^{\alpha}$ and contricity
of the traceless $\hat{Q}_{\alpha\beta\mu}$, and $\Gamma_{\;\;\;\beta\mu}^{\alpha}$
is the Levi-Civita connection. Computing the form of the remaining
tensors, the full invariant geometry becomes
\begin{eqnarray}
\mathfrak{T}_{\;\;\;\mu\nu}^{\alpha} & = & T_{\;\;\;\mu\nu}^{\alpha}\nonumber \\
\mathfrak{Q}_{\alpha\beta\mu} & = & \hat{Q}_{\alpha\beta\mu}=Q_{\alpha\beta\mu}-\frac{1}{n}g_{\alpha\beta}Q_{\;\;\;\alpha\mu}^{\alpha}\nonumber \\
\mathfrak{R}_{\;\;\;\beta\mu\nu}^{\alpha} & = & C_{\;\;\;\beta\mu\nu}^{\alpha}+2\Delta_{\mu\beta}^{\alpha\rho}\left(\mathcal{R}_{\rho\nu}+W_{\left(\rho;\nu\right)}+W_{\rho}W_{\nu}-\frac{1}{2}W^{2}g_{\rho\nu}\right)\nonumber \\
 &  & -2\Delta_{\nu\beta}^{\alpha\rho}\left(\mathcal{R}_{\rho\mu}+W_{\left(\rho;\mu\right)}+W_{\rho}W_{\mu}-\frac{1}{2}W^{2}g_{\rho\mu}\right)-2\left(\Delta_{\mu\beta}^{\alpha\rho}\Omega_{\nu\rho}-\Delta_{\nu\beta}^{\alpha\rho}\Omega_{\mu\rho}\right)\label{Tensors of Weyl geometry}
\end{eqnarray}
where $C_{\;\;\;\beta\mu\nu}^{\alpha}$ is the conformal curvature,
$\Delta_{\nu\beta}^{\alpha\rho}\equiv\frac{1}{2}\left(\delta_{\nu}^{\alpha}\delta_{\beta}^{\rho}-g^{\alpha\rho}g_{\beta\nu}\right)$
and $\mathcal{R}_{\alpha\beta}\equiv-\frac{1}{n-2}\left(R_{\alpha\beta}-\frac{1}{2\left(n-1\right)}g_{\alpha\beta}R\right)$
is the Schouten tensor \cite{Schouten}. In an integrable Weyl geometry
the dilatational field strength $\Omega_{\alpha\beta}\equiv W_{\alpha,\beta}-W_{\beta,\alpha}$
vanishes so the only dependence on the Weyl vector comes in the additions
to the Schouten tensor. These additions vanish in the Riemannian gauge,
$\tilde{W}_{a}=0$.

\subsection{Conformal versus projective \label{subsec:Conformal-versus-projective}}

In the last two Subsections we have seen that both the projectively
and conformally invariant fields may be written by including certain
combinations of the Weyl vector with the original fields. This means
that the Riemannian gauge $W_{\alpha}=0$ gives the same curvature,
torsion, and traceless nonmetricity in both the projective and conformal
cases.
\[
\begin{array}{ccccc}
\mathfrak{R}_{\;\;\;bcd}^{a} & = & P_{\;\;\;\beta\mu\nu}^{\alpha} & = & R_{\;\;\;\beta\mu\nu}^{\alpha}\\
\mathfrak{T}_{\;\;\;\mu\nu}^{\alpha} & = & \mathcal{T}_{\;\;\;\beta\mu}^{\alpha} & = & T_{\;\;\;\beta\mu}^{\alpha}\\
\mathfrak{Q}_{\alpha\beta\mu} & = & \mathcal{Q}_{\alpha\beta\mu} & = & \hat{Q}_{\alpha\beta\mu}
\end{array}
\]
In other gauges, the expressions differ only by the scaling of the
Weyl vector and the resulting form of the curvatures.

In \cite{Matveev2} the existence of such a gauge is proved by showing
gauge equivalence between Weyl symmetry of the light cones and projective
symmetry of the timelike geodesics. Here, we showed explicitly that
equivalence occurs in the Riemannian gauge, for all values of $a$
for the projective transformation, and without appealing to the light
cone structure.

\subsubsection{Projective and conformal Ricci flatness}

Further evidence of the close relationship between projective and
conformal symmetries is seen by comparing the conditions for a spacetime
to be either projectively or conformally Ricci flat. Though the conditions
are well-known, we provide a concise derivation for the projective
case. See \cite{Wheeler2018a} for a similar derivation of the conformal
condition.

Using Eq.(\ref{Projective change in R}) and assuming vanishing trace
of the torsion, the condition for projective Ricci flatness follows
easily by writing the Ricci tensor as a 1-form, and requiring the
transformed Ricci tensor $\tilde{R}_{\beta\nu}\mathbf{d}x^{\nu}$
to vanish. From Eq.(\ref{Projective change in R}) we have
\begin{eqnarray*}
\mathbf{d}\xi_{,\beta} & = & \frac{1}{\left(1-a\right)\left(n-1\right)}\mathbf{R}_{\beta}+\xi_{,\alpha}\boldsymbol{\Sigma}_{\;\;\;\beta}^{\alpha}+\left(1-a\right)\xi_{,\beta}\mathbf{d}\xi
\end{eqnarray*}
The integrability condition $\mathbf{d}^{2}\xi_{,\beta}\equiv0$ leads
directly to
\begin{eqnarray*}
0 & = & \mathbf{D}\mathbf{R}_{\beta}+\left(1-a\right)\left(n-1\right)\xi_{,\alpha}\mathbf{R}_{\;\;\;\beta}^{\alpha}+\left(1-a\right)\mathbf{R}_{\beta}\land\mathbf{d}\xi
\end{eqnarray*}
Rescaling $\phi_{,\sigma}=\left(n-1\right)\left(1-a\right)\xi_{,\sigma}$
we have
\begin{equation}
0=\mathbf{D}\mathbf{R}_{\beta}+\phi_{,\alpha}\mathbf{R}_{\;\;\;\beta}^{\alpha}+\frac{1}{n-1}\mathbf{R}_{\beta}\land\mathbf{d}\phi\label{Projective Ricci flatness}
\end{equation}
This is the condition for the existence of a projective tranformation
leading to the vacuum Einstein equation.

It is interesting to observe that if we choose $a=1$ then integrability
requires the curl of the Ricci tensor to vanish. At the same time,
contractions of the Bianchi identity imply the vanishing divergences
\begin{eqnarray*}
R_{\;\;\;bde;c}^{c} & = & 0\\
R_{\;\;\;d;c}^{c} & = & 0
\end{eqnarray*}
With both vanishing curl and divergence, the Ricci tensor satisfies
the covariant wave equation
\[
\left(^{*}\mathbf{D}^{*}\mathbf{D}+\mathbf{D}^{*}\mathbf{D}^{*}\right)\mathbf{R}_{a}=\square R_{ab}=0
\]
Therefore, in the Thomas gauge, the wave equation of the Ricci tensor
implies projective Ricci flatness.

There is strong similarity between the projective condition (\ref{Projective Ricci flatness})
and the well-known conformal condition for Ricci flatness,
\begin{eqnarray*}
0 & = & \mathbf{D}\mathbf{R}_{\beta}+\phi_{,\sigma}\mathbf{C}_{\;\;\;\beta}^{\sigma}
\end{eqnarray*}
If we write the full curvature tensor in Eq.(\ref{Projective Ricci flatness})
in terms of the Weyl curvature and the appropriate Ricci terms,
\begin{eqnarray*}
\mathbf{R}_{\;\;\;\beta}^{\sigma} & = & \mathbf{C}_{\;\;\;\beta}^{\sigma}-\frac{2}{n-2}\Delta_{\nu\beta}^{\sigma\alpha}\left(\mathbf{R}_{\alpha}-\frac{1}{2\left(n-1\right)}Rg_{\alpha\mu}\mathbf{d}x^{\mu}\right)\land\mathbf{d}x^{\nu}
\end{eqnarray*}
then the difference of the conditions, $\boldsymbol{\Delta}_{\beta}=\left(\mathbf{D}\mathbf{R}_{\beta}+\phi_{,\sigma}\mathbf{C}_{\;\;\;\beta}^{\sigma}\right)-\left(\mathbf{D}\mathbf{R}_{\beta}+\phi_{,\sigma}\mathbf{R}_{\;\;\;\beta}^{\sigma}+\frac{1}{n-1}\mathbf{R}_{\beta}\wedge\mathbf{d}\phi\right)$
depends on the Ricci tensor alone.
\begin{eqnarray*}
\boldsymbol{\Delta}^{\beta} & = & \frac{1}{\left(n-1\right)\left(n-2\right)}\left(\mathbf{R}^{\beta}\wedge\mathbf{d}\phi+\left(R\mathbf{d}\phi-\left(n-1\right)\phi^{,\alpha}\mathbf{R}_{\alpha}\right)\land\mathbf{d}x^{\beta}\right)
\end{eqnarray*}
Clearly, $\boldsymbol{\Delta}_{\beta}=0$ if the Ricci tensor vanishes.
It follows that if either of the conditions for Ricci flatness is
satisfied, then there exists a gauge where both conditions are satisfied.
As Stachel \cite{Stachel} puts it, ``If the homogeneous Einstein
field equations hold, the difference between these curvature tensors
is purely conceptual.'' 

\section{Existence of $\xi$ in a suitable domain of dependence\label{sec:Existence-of-}}

We have shown that both conformal and projective symmetry rely on
the existence of a suitable function $\xi$. However, while it is
always possible to reparameterize a given curve $x^{\alpha}\left(\lambda\right)$
by defining any differentiable function $\sigma\left(\lambda\right):\mathbb{R}^{1}\rightarrow\mathbb{R}^{1}$,
it is not always true that this can be extended to a function $\sigma\left(x^{\alpha}\right):\mathbb{R}^{n}\rightarrow\mathbb{R}^{1}$
in a region. Any reparameterization must maintain the progression
of the curve, so $\sigma$ must be monotonic on all affected curves.
This is true for conformal transformations as well, since we need
a positive conformal factor, and $e^{\xi}=\frac{d\sigma}{d\lambda}>0$.
Further, we showed in Subsection (\ref{subsec:Properties-of}) that
in the absence of curvature singularities $\sigma\left(x^{\alpha}\right)$
must have at least continuous second derivatives.

We begin with a concrete construction of a satisfactory function under
suitable conditions, then discuss two further issues surrounding extensions:
(1) whether measurements might distinguish between projective and
conformal symmetries of spacetime, and (2) topological obstructions
to the existence of $\sigma\left(x^{\alpha}\right)$.

\subsection{Example \label{subsec:Example}}

For an example of an allowed function, suppose we have a region $\mathscr{R}\subset\mathcal{M}^{3}\times\mathbb{R}$
with a single timelike direction, foliated by like-oriented timelike
geodesics, with spacelike cross-section $\mathcal{M}^{3}$. Let $\tau\left(x^{\alpha}\right)=\tau\left(\tau,\mathbf{x}\right)$
be the proper time along each curve of the foliation, one through
each $\mathbf{x}\in\mathcal{M}^{3}$. This associates a value of $\tau$
with every point $x^{\alpha}$ in the region, providing one suitable
parameterization for all of $\mathscr{R}$. Forgetting that $\tau$
is proper time on the foliation of geodesics, take $\tau$ as a timelike
coordinate on the region. The function $\lambda=\tau\left(x^{\alpha}\right)$
is monotonic along all timelike curves contained in the region and
may be used to parameterize all timelike curves in the region. Then
a regional projective or conformal transformation $\sigma\left(\lambda\left(x^{\alpha}\right)\right)$
is the composition of any monotonic $C^{2}$ function $\sigma\left(\lambda\right)$
with $\lambda\left(x^{\alpha}\right)$. 

\subsection{Projective versus conformal structures}

We have shown that invariance under regional reparameterizations of
geodesics can be incorporated into the spacetime geometry by introducing
either projectively or conformally invariant connections. In keeping
with the Ehlers, Pirani, Schild program we now consider whether measuring
many geodesics allows us to distinguish between these two different
symmetries. 

First, consider whether different transformations lead to different
restrictions on the function $\xi$. For writing a projective connection
we need a $C^{2}$ regional reparameterization function $\sigma\left(x^{\alpha}\right)$,
leading to
\[
\tilde{\Gamma}_{\left(a\right)\;\mu\nu}^{\;\alpha}=\Gamma_{\;\;\;\mu\nu}^{\alpha}+a\delta_{\mu}^{\alpha}\xi_{,\nu}+\left(1-a\right)\delta_{\nu}^{\alpha}\xi_{,\mu}
\]
where $\xi=\ln\frac{d\sigma\left(x\right)}{d\lambda}$ while conformal
transformation depends on the \emph{same} function
\[
\tilde{g}_{\alpha\beta}=e^{2\xi}g_{\alpha\beta}
\]
where $\xi$ satisfies the same differentiability condition.

We show that \emph{any} $\mathcal{C}^{1}$ function $\xi$ leads to
an acceptable reparameterization $\sigma\left(\lambda\right)$. Choose
any $\mathcal{C}^{1}$ function $\xi\left(x^{\alpha}\right)$. To
find a reparameterization $\sigma\left(\lambda\right)$ corresponding
to $\xi$, we must solve
\[
\frac{d\sigma}{d\lambda}=e^{\xi}
\]
for all geodesics in the region. There is no limitation due to the
exponential because because $\sigma$ must be monotonic. To see that
any $\mathcal{C}^{1}$ function $\xi$ allows a solution, return to
our congruence of timelike geodesics in (\ref{subsec:Example}), each
parameterized by its proper time. Along this congruence define
\[
\sigma\left(\tau\right)=\intop_{C;\,0}^{\tau}e^{\xi\left(\tau'\right)}d\tau'
\]
where $\xi\left(\tau'\right)$ is the value of $\xi\left(x^{\alpha}\right)$
along the geodesic $C$ at position $\tau'$. This gives a value for
$\sigma\left(x^{\alpha}\right)$ at every point within the foliation
from our initial surface, and satisfies
\[
\frac{d\sigma}{dt}=\frac{d\tau}{dt}\frac{d\sigma}{d\tau}=e^{\xi}
\]
where $\frac{d\tau}{dt}=1$, since the time parameter $t$ on all
timelike geodesics is taken as $\tau$ on the congruence. Hence, there
is no restriction on $\xi$ that would allow us to distinguish projective
from conformal.

Next, we examine whether measurements of more than individual geodesics
can separate projective from conformal transformations. This cannot
be the case, since we have found invariant forms of the tensors describing
the geometry. Remembering that a symmetry represents the things that
we can \emph{not} measure, if there exists a choice for $\xi$ such
that the projective and conformal geodesics, torsions, nonmetricities,
and curvatures are all equal, then they are not distinguishable. Other
gauges are merely distinct ways to represent the same theory. We have
shown in Subsection \ref{subsec:Conformal-versus-projective} that
such a gauge exists for any $a$, and that in that gauge the curvature,
torsion, and nonmetricity take their usual forms.

\subsection{Topological questions}

Based on simple examples, we argue for necessary and sufficient restrictions:
a region foliated by order isomorphic, totally ordered, $C^{2}$ timelike
curves \cite{Wolfram,Wolfram 2}. We begin with some examples, then
provide rigorous definitions.

To simplify the discussion, we consider torsion-free spacetime $\left(\mathcal{M},g\right)$.
Suppose the metric $g$ has two timelike directions spanning a region
of a 2-dimensional submanifold $\mathcal{M}^{2}$. Then the set of
all timelike geodesics through any fixed point may be distinguished
by an angle characterizing the initial direction in the plane. As
this angle increases by $\pi$ the direction must abruptly change.
Any regional reparameterization will not be differentiable at the
discontinuity. This form of the problem resolves if we limit $g$
to a single timelike dimension, since then we cannot construct such
a rotational continuum with tangent timelike curves. However, there
is a further difficulty. 

With one dimension, we might suppose that a region with $\mathcal{M}^{3}\times\left[a,b\right]$
topology, with $\left[a,b\right]\subset\mathbb{R}$ timelike and $\mathcal{M}^{3}$
spacelike would work, but this is still not strong enough\footnote{A Möbius strip is not a counterexample, since any open subset of the
strip has a common time direction. We need a region with a discontinuity.}. Common time orientation is necessary, and this is a stronger condition.
For example, consider $\mathbb{R}^{2}$ with Minkowski metric and
Cartesian coordinates $\left(x,t\right)$ for $x\geq0$ and $\left(x,-t\right)$
for $x<0$. Although we have only a single time coordinate, declare
the straight line geodesics $\left(x_{0},\pm t\right)$ to have orientation
given by tangents $u^{\alpha}=\frac{dx^{\alpha}}{d\left(\pm t\right)}$.
This choice allows no regional reparameterization containing any point
of the curve $\left(0,t\right)$. We again have a discontinuity.

The required property is time orientability on the region. To define
time orientability, we first separate past and future regions with
an equivalence relation, $x^{\alpha}\cong y^{\alpha}$ iff $\left\langle x,y\right\rangle =g_{\alpha\beta}x^{\alpha}y^{\beta}<0$.
In spacetime there are exactly two regions since any two future oriented
vectors are equivalent, and any two past oriented vectors are equivalent,
while one vector from the future and one from the past are inequivalent.
A region is time orientable if a consistent assignment of future direction
exists over the entire region.

The equivalence relation is conformally invariant, since $g_{\alpha\beta}x^{\alpha}y^{\beta}<0$
iff $e^{2\xi}g_{\alpha\beta}x^{\alpha}y^{\beta}<0$. Conversely, the
conformal invariance means that we can reparameterize using any bounded
function $\xi$, since $\left(e^{2\xi}g_{\alpha\beta}\right)x^{\alpha}y^{\beta}=g_{\alpha\beta}\left(e^{\xi}x^{\alpha}\right)\left(e^{\xi}y^{\beta}\right)$.

Notice that time orientability requires restriction to a single time
direction. For spacetimes the signature is $\left(n-1,1\right)$ and
the equivalence relation takes the form $-x^{0}y^{0}+\mathbf{x}\cdot\mathbf{y}<0$
where negative norm implies $x^{0}>\left|\mathbf{x}\right|$ and $y^{0}>\left|\mathbf{y}\right|$
. We also have $\mathbf{x}\cdot\mathbf{y}<\left|\mathbf{x}\right|\cdot\left|\mathbf{y}\right|$
so $-x^{0}y^{0}+\left|\mathbf{x}\right|\cdot\left|\mathbf{y}\right|<0$.
However, if we have two or more timelike directions then $x^{\alpha}=\left(\mathbf{t},\mathbf{x}\right)$
so that timelike vectors satisfy $-\mathbf{t}\cdot\mathbf{t}+\mathbf{x}\cdot\mathbf{x}<0$.
Now if we try to form equivalence classes we have
\begin{align*}
-\mathbf{t}_{1}\cdot\mathbf{t}_{2}+\mathbf{x}\cdot\mathbf{y} & <0\\
-\left|\mathbf{t}_{1}\right|\cdot\left|\mathbf{t}_{2}\right|\cos\theta+\left|\mathbf{x}\right|\left|\mathbf{y}\right|\cos\phi & <0
\end{align*}
As $\theta$ and $\phi$ vary we can get both signs, regardless of
the magnitudes. Therefore, time orientability exists only in Lorentzian
spacetimes. Similarly, spatial orientability requires $\left(p,q\right)=\left(1,q\right)$.
Full orientability, for both space and time directions can occur in
2-dimensions and only for signature $\left(1,1\right)$, as possessed
by string world sheets. This has some implications for tachyons, since
these follow spacelike geodesic trajectories. Such spacelike geodesics
do not permit a formulation invariant under spacelike reparameterizations
of curves.

\section{Linear conformal connection \label{sec:Linear-representation-of}}

The second axis of the Ehlers, Pirani, Schild program \cite{EPS original,EPS}
is the connection inferred from lightlike geodesics. They define a
linear connection on 4-dimensional spacetime in terms of a metric
tensor density. This leads to an integrable Weyl connection. In subsequent
works \cite{Matveev1,Matveev2} Weyl structure is defined and assumed
for the conformal connection.

Having shown that regional reparameterization of timelike geodesics
also leads to a conformal geometry, it might seem that we are done.
However, since the 4-dimensional representation of special conformal
transformations is nonlinear, the implicit choice in \cite{EPS original,EPS,Matveev1,Matveev2}
to use linear connections in 4-dimensions limits the full conformal
symmetry to a Weyl connection. The lowest dimension for a real, linear
representation of the conformal group $\left(SO\left(4,2\right)\right)$
is six. We have shown that the regional reparameterization of timelike
geodesics may be accomplished by transformations preserving the conformal
class of metrics $\left\{ e^{2\xi}g_{\alpha\beta}\right\} $, without
restriction on $\xi$. All conformal transformations satisfy this,
with the same condition applying to null geodesics. We therefore need
no comparison of null and timelike structures. Instead, we develop
a real linear representation of all conformal transformations. In
the next Section, we use these to write and employ a linear conformal
connection on spacetime.

There has been extensive study of gauge theories of gravity with conformal
symmetry. We give a brief overview of some of the previous treatments
of conformal gauging.

Prior to 1977, conformal gauging incorporated only Lorentz transformations
and dilatations. Therefore, when Ehlers, Pirani, and Schild discussed
the light cone symmetry, the Weyl group was regarded as the unique
gauge theory of the conformal group. Various gauge theories of gravity
included this reduced symmetry. In order to remain as close as possible
to Einstein's theory, Deser \cite{Deser} coupled a massless Lorentz
scalar field $\phi(x)$ (dilaton) of compensating conformal weight
$w=-1$ to gravitation through the manifestly scale-invariant quantity
$\frac{1}{6}\phi^{2}R$. Later, Dirac \cite{Dirac}, trying to accommodate
the Large Numbers Hypothesis, similarly modified Weyl's free Lagrangian
by replacing all $R^{2}$-type terms by $\phi^{2}R$. This method
gave rise to various theories involving the ``generalized'' Einstein
equations \cite{Freund,Bramson,Omote,Kasuya,Bicknell}. They were
shown to reduce to general relativity when expressed in a particular
gauge \cite{Freund,Bramson}, equivalent to an integrable Weyl geometry.

In 1977, the full conformal group was gauged by Crispim-Romao, Ferber
and Freund \cite{Freund2,Romao+Ferber+Freund} and independently by
Kaku, Townsend and Van Nieuwenhuizen \cite{Kaku+Townsend,Kaku+Townsend2},
by treating the conformal group in four dimensions in much the same
way as Poincaré gauging, handling the dilatations and special conformal
transformations as generators of additional symmetries. A comprehensive
study of the gauged conformal group and its subgroups was subsequently
given by Ivanov and Niederle \cite{Ivanov1,Ivanov2}. Using a conformally
invariant $4$-dimensional action quadratic in the conformal curvatures
and the assumption of vanishing torsion, it is found that the gauge
fields associated with special conformal transformations are algebraically
removable. The action reduces to the conformally invariant, torsion-free
Weyl theory of gravity based on the square of the conformal curvature\footnote{Note the difference between Weyl \emph{geometry}--a geometry including
dilatations in addition to Lorentz or Poincarè symmetry--and Weyl
\emph{gravity}. Weyl gravity is a theory of gravity with action quadratic
in the conformal curvature and posessing full conformal symmetry.}. This auxiliary nature of the special conformal gauge field has been
shown to follow for any $4$-dimensional action quadratic in the curvatures
\cite{Wheeler1991}, and we extend this result below by showing that
the result follows directly from the structure equations, independently
of any action. Generically, the action reduces to the a linear combination
of the square of the conformal curvature and the square of the curl
of the Weyl vector. When all conformal gauge fields (not just the
metric as in Weyl gravity) are varied, auxiliary conformal gravity
reduces to scale-invariant general relativity (i.e., Ricci-flat, integrable
Weyl geometry) \cite{Wheeler 2014}.

We conclude this Section with a modified presentation of this gauge
theory of gravity, proving the stronger result for the auxiliary character
of the gauge field of special conformal transformations. We then vary
the simplest quadratic action and describe how it leads to integrable
Weyl geometry as in \cite{Wheeler 2014}.

\subsection{Spacetime with a conformal connection\label{sec:Spacetime-with-a}}

Once it is shown that the conformal group has a representation as
$SO\left(4,2\right)$ we may simply write the $SO\left(4,2\right)$
Cartan equations.
\begin{eqnarray*}
\boldsymbol{\Omega}_{\;\;\;B}^{A} & = & \mathbf{d}\boldsymbol{\Sigma}_{\;\;\;B}^{A}-\boldsymbol{\Sigma}_{\;\;\;B}^{C}\wedge\boldsymbol{\Sigma}_{\;\;\;C}^{A}
\end{eqnarray*}
where $A,B,\ldots=0,\ldots5$, with $\boldsymbol{\Sigma}_{\;\;\;B}^{A}$
the $SO\left(4,2\right)$ Cartan connection.

Taking the quotient of the conformal group by its inhomogeneous Weyl
subgroup leads to a 4-dimensional homogeneous manifold $\mathcal{M}_{0}^{4}$.
Defining a projection from the cosets to the quotient manifold then
yields a principal fiber bundle with inhomogeneous Weyl symmetry.
Introducing horizontal curvatures $\boldsymbol{\Omega}_{\;\;\;B}^{A}=\frac{1}{2}\Omega_{\;\;\;Bcd}^{A}\mathbf{e}^{c}\wedge\mathbf{e}^{d}$,
preserves the fiber bundle structure, now over any choice of manifold
$\mathcal{M}^{4}$, with curvature.

Having a real, linear representation of conformal transformations,
a linear conformal connection on $\mathcal{M}^{4}$ is given by
\begin{eqnarray*}
\boldsymbol{\Sigma}_{\;\;\;B}^{A} & = & \Sigma_{\;\;\;B\alpha}^{A}\mathbf{d}x^{\alpha}=\Sigma_{\;\;\;Ba}^{A}\mathbf{e}^{a}
\end{eqnarray*}
where $\mathbf{e}^{a}$ is an orthonormal basis form in 4-dimensions,
but the upper case Latin indices run over the linear representation,
$A,B=0,1,2,3,4,5$. We may decompose $\boldsymbol{\Sigma}_{\;\;\;B}^{A}$
into Lorentz and dilatationally invariant parts
\begin{equation}
\boldsymbol{\Sigma}_{\;\;\;B}^{A}=\left(\boldsymbol{\Sigma}_{\;\;\;b}^{a},\boldsymbol{\Sigma}_{\;\;\;4}^{a},\boldsymbol{\Sigma}_{\;\;\;a}^{4},\boldsymbol{\Sigma}_{\;\;\;4}^{4}\right)=:\left(\boldsymbol{\omega}_{\;\;\;b}^{a},\mathbf{e}^{a},\mathbf{f}_{a},\boldsymbol{\omega}\right)\label{Weyl invariant parts}
\end{equation}
where $\boldsymbol{\omega}_{\;\;\;b}^{a}$ is the 4-dimensional spin
connection, $\mathbf{e}^{a},\mathbf{f}_{a}$ are the gauge fields
of translations and special conformal transformations (co-translations),
and $\boldsymbol{\omega}$ is the Weyl vector. The 1-forms $\mathbf{e}^{a}=e_{\mu}^{\;\;\;a}\mathbf{d}x^{\mu}$
form an orthonormal basis. This leads to

\begin{eqnarray}
\boldsymbol{\Omega}_{\;\;\;b}^{a} & = & \mathbf{d}\boldsymbol{\omega}_{\;\;\;b}^{a}-\boldsymbol{\omega}_{\;\;\;b}^{c}\wedge\boldsymbol{\omega}_{\;\;\;c}^{a}+\left(\delta_{d}^{a}\delta_{b}^{c}-\eta^{ac}\eta_{bd}\right)\mathbf{e}^{d}\land\mathbf{f}_{c}\nonumber \\
\mathbf{T}^{a} & = & \mathbf{d}\mathbf{e}^{a}-\mathbf{e}^{b}\wedge\boldsymbol{\omega}_{\;\;\;b}^{a}-\boldsymbol{\omega}\wedge\mathbf{e}^{a}\nonumber \\
\mathbf{S}_{a} & = & \mathbf{d}\mathbf{f}_{a}-\boldsymbol{\omega}_{\;\;\;a}^{b}\wedge\mathbf{f}_{b}-\mathbf{f}_{a}\wedge\boldsymbol{\omega}\nonumber \\
\boldsymbol{\Theta} & = & \mathbf{d}\boldsymbol{\omega}-\mathbf{e}^{c}\land\mathbf{f}_{c}\label{SO(4,2) Cartan equations}
\end{eqnarray}
where an additional exterior derivative gives the integrability conditions
(generalized Bianchi identities). Note that the components of the
curvature $\boldsymbol{\Omega}_{\;\;\;B}^{A}=\left(\boldsymbol{\Omega}_{\;\;\;b}^{a},\mathbf{T}^{a},\mathbf{S}_{a},\boldsymbol{\Theta}\right)$
in (\ref{SO(4,2) Cartan equations}) mix under conformal transformation.

\subsubsection{Reduction using gauge symmetry}

We can make use of conformal gauge transformations and other considerations
to simplify the expressions for the various components of the curvature.

First, notice that under full conformal transformation the Lorentz
part of the curvature tensor transforms as
\begin{eqnarray*}
\tilde{\boldsymbol{\Omega}}_{\;\;\;b}^{a} & = & \Lambda_{\;\;\;c}^{a}\boldsymbol{\Omega}_{\;\;\;b}^{a}\bar{\Lambda}_{\;\;\;b}^{d}\\
 &  & +\Lambda_{\;c}^{a}\mathbf{T}^{c}\bar{\Lambda}_{b}+\eta^{ac}\Lambda_{c}\mathbf{T}^{e}\eta_{ed}\bar{\Lambda}_{\;\;\;b}^{d}\\
 &  & +\Lambda^{a}\mathbf{S}_{c}\bar{\Lambda}_{\;\;\;b}^{c}+\Lambda_{\;\;\;d}^{a}\eta^{de}\mathbf{S}_{e}\eta_{bc}\bar{\Lambda}^{c}
\end{eqnarray*}
This is independent of dilatations. For $\boldsymbol{\Omega}_{\;\;\;b}^{a}$
to remain invariant under dilatations it must be one of the three
conformally invariant tensors, i.e., the Bach tensor, the Cotton tensor,
or the Weyl curvature. Its rank and full dependence on the Riemann
tensor restrict this to the Weyl curvature, $\mathbf{C}_{\;\;\;b}^{a}$.

Next, acting on the connection with the full conformal group changes
the connection forms inhomogeneously as follows. 

\begin{eqnarray*}
\tilde{\boldsymbol{\omega}}_{\;b}^{a} & = & \Lambda_{\;c}^{a}\boldsymbol{\omega}_{\;d}^{c}\bar{\Lambda}_{vb}^{d}-\mathbf{d}\Lambda_{\;\;\;c}^{a}\:\bar{\Lambda}_{\;\;\;b}^{c}+\Lambda_{\;c}^{a}\mathbf{e}^{c}\bar{\Lambda}_{b}\\
 &  & +\eta^{ac}\Lambda_{c}\mathbf{e}^{e}\eta_{ed}\bar{\Lambda}_{\;\;\;b}^{d}+\Lambda^{a}\mathbf{f}_{c}\bar{\Lambda}_{\;\;\;b}^{c}+\Lambda_{\;\;\;d}^{a}\eta^{de}\mathbf{f}_{e}\eta_{bc}\bar{\Lambda}^{c}\\
\tilde{\mathbf{e}}^{a} & = & \Lambda_{\;\;\;c}^{a}\mathbf{e}^{c}\bar{\Lambda}+\Lambda_{\;\;\;c}^{a}\boldsymbol{\omega}_{\;\;\;d}^{c}\bar{\Lambda}^{d}+\Lambda^{a}\mathbf{e}^{c}\bar{\Lambda}_{c}\\
 &  & +\Lambda^{a}\boldsymbol{\omega}\bar{\Lambda}-\Lambda_{\;\;\;c}^{a}\mathbf{d}\Lambda^{c}-\Lambda^{a}\mathbf{d}\Lambda\\
\tilde{\mathbf{f}}_{a} & = & \Lambda\mathbf{f}_{c}\bar{\Lambda}_{\;\;\;a}^{c}+\Lambda_{c}\boldsymbol{\omega}_{\;\;\;d}^{c}\bar{\Lambda}_{\;\;\;a}^{d}+\Lambda^{c}\mathbf{f}_{c}\bar{\Lambda}_{a}\\
 &  & +\Lambda\boldsymbol{\omega}\bar{\Lambda}_{a}-\mathbf{d}\Lambda_{c}\:\bar{\Lambda}_{\;\;\;a}^{c}-\mathbf{d}\Lambda\:\bar{\Lambda}_{a}\\
\tilde{\boldsymbol{\omega}} & = & \boldsymbol{\omega}+\Lambda_{c}\mathbf{e}^{c}\bar{\Lambda}+\Lambda\boldsymbol{\omega}_{c}\bar{\Lambda}^{c}-\mathbf{d}\Lambda\:\bar{\Lambda}
\end{eqnarray*}
Curving the base manifold spanned by $\mathbf{e}^{a}$ removes the
translational symmetry $\Lambda^{a}$. Temporarily holding the dilatational
transformation fixed $\left(\Lambda=1\right)$ only the co-translations
affect the Weyl vector, $\tilde{\boldsymbol{\omega}}=\boldsymbol{\omega}+\Lambda_{c}\mathbf{e}^{c}$.
Since $\Lambda_{c}\mathbf{e}^{c}$ is an arbitrary 1-form we may choose
it to cancel $\boldsymbol{\omega}$, so in this gauge the Weyl vector
vanishes. This exhausts our co-translational freedom. 

Now, restoring dilatations while dropping $\boldsymbol{\omega}$ and
transformations depending on $\Lambda^{a}$ or $\Lambda_{c}$, we
have
\begin{eqnarray*}
\tilde{\boldsymbol{\omega}}_{\;b}^{a} & = & \Lambda_{\;c}^{a}\boldsymbol{\omega}_{\;d}^{c}\bar{\Lambda}_{\;\;\;b}^{d}-\mathbf{d}\Lambda_{\;\;\;c}^{a}\:\bar{\Lambda}_{\;\;\;b}^{c}\\
\tilde{\mathbf{e}}^{a} & = & \Lambda_{\;\;\;c}^{a}\mathbf{e}^{c}\bar{\Lambda}\\
\tilde{\mathbf{f}}_{a} & = & \Lambda\mathbf{f}_{c}\bar{\Lambda}_{\;\;\;a}^{c}\\
\tilde{\boldsymbol{\omega}} & = & -\mathbf{d}\Lambda\:\bar{\Lambda}
\end{eqnarray*}
These are the gauge transformations of an integrable Weyl geometry,
with an extra tensor field $\mathbf{f}_{a}=\mathsf{f}_{ab}\mathbf{e}^{b}$.

\subsubsection{The Weyl curvature}

With these gauge choices we may write the Cartan equations in the
simpler form
\begin{eqnarray}
\boldsymbol{\Omega}_{\;\;\;b}^{a} & = & \mathbf{d}\boldsymbol{\omega}_{\;\;\;b}^{a}-\boldsymbol{\omega}_{\;\;\;b}^{c}\wedge\boldsymbol{\omega}_{\;\;\;c}^{a}+\left(\delta_{d}^{a}\delta_{b}^{c}-\eta^{ac}\eta_{bd}\right)\mathbf{e}^{d}\land\mathbf{f}_{c}\label{Lorentz curvature}\\
\mathbf{T}^{a} & = & \mathbf{d}\mathbf{e}^{a}-\mathbf{e}^{b}\wedge\boldsymbol{\omega}_{\;\;\;b}^{a}=\mathbf{D}\mathbf{e}^{a}\label{Solder form}\\
\mathbf{S}_{a} & = & \mathbf{d}\mathbf{f}_{a}-\boldsymbol{\omega}_{\;\;\;a}^{b}\wedge\mathbf{f}_{b}=\mathbf{D}\mathbf{f}_{a}\label{Co-solder form}\\
\boldsymbol{\Theta} & = & -\mathbf{e}^{c}\land\mathbf{f}_{c}\label{Dilatation}
\end{eqnarray}
Equation (\ref{Solder form}) for the solder form may be solved for
the spin connection, which is also that of an integrable Weyl geometry
with torsion, where (since we are in the Riemann gauge) $\mathbf{R}_{\;\;\;b}^{a}=\mathbf{d}\boldsymbol{\omega}_{\;\;\;b}^{a}-\boldsymbol{\omega}_{\;\;\;b}^{c}\wedge\boldsymbol{\omega}_{\;\;\;c}^{a}$
is Riemann curvature tensor, plus the usual contorsion terms.

The Weyl curvature and Riemann curvature are related by
\begin{eqnarray}
\mathbf{C}_{\;\;\;b}^{a} & = & \mathbf{R}_{\;\;\;b}^{a}+\left(\delta_{d}^{a}\delta_{b}^{c}-\eta^{ac}\eta_{bd}\right)\mathbf{e}^{d}\land\boldsymbol{\mathcal{R}}_{c}\label{Weyl curvature}
\end{eqnarray}
where $\boldsymbol{\mathcal{R}}_{c}$ is the Schouten tensor 1-form,
$\boldsymbol{\mathcal{R}}_{c}=-\frac{1}{n-2}\left(R_{ab}-\frac{1}{2\left(n-1\right)}\eta_{ab}R\right)\mathbf{e}^{b}$.
Subtracting Eq.(\ref{Weyl curvature}) from Eq.(\ref{Lorentz curvature})
leaves
\begin{eqnarray*}
\left(\delta_{d}^{a}\delta_{b}^{c}-\eta^{ac}\eta_{bd}\right)\mathbf{e}^{d}\land\left(\mathbf{f}_{c}-\boldsymbol{\mathcal{R}}_{c}\right) & = & 0
\end{eqnarray*}
Expanding this into components with $\mathbf{f}_{c}-\boldsymbol{\mathcal{R}}_{c}=\left(\mathsf{f}_{ce}-\mathcal{R}_{ce}\right)\mathbf{e}^{e}$
and taking a contraction on $ad$, gives $\left(n-2\right)\left(\mathsf{f}_{be}-\mathcal{R}_{be}\right)+\eta_{be}\left(\mathsf{f}_{\;\;\;a}^{a}-\mathcal{R}_{\;\;\;a}^{a}\right)=0$.
The $be$ trace shows that $\mathsf{f}_{\;\;\;a}^{a}-\mathcal{R}_{\;\;\;a}^{a}=0$
, so for $n\neq2$ we are left with
\begin{eqnarray}
\mathbf{f}_{b} & = & \boldsymbol{\mathcal{R}}_{b}\label{f is auxiliary}
\end{eqnarray}
showing that $\boldsymbol{\Omega}_{\;\;\;b}^{a}=\mathbf{C}_{\;\;\;b}^{a}$
and $\mathbf{S}_{a}=\mathbf{D}\boldsymbol{\mathcal{R}}_{a}$. Furthermore
$\mathbf{e}^{b}\wedge\mathbf{f}_{b}=\mathbf{e}^{b}\wedge\boldsymbol{\mathcal{R}}_{b}=0$
by the symmetry of $\mathcal{R}_{ab}$ so that Eq.(\ref{Dilatation})
reduces to vanishing dilatation, $\boldsymbol{\Theta}=0$. Since Eq.(\ref{f is auxiliary})
follows directly from the Cartan structure equations this is a stronger
result than proved in \cite{Wheeler1991}. The only unreduced structure
equation is $\mathbf{D}\mathbf{e}^{a}=\mathbf{T}^{a}$. The torsion-free
gravity theory built from these elements is essentially unique.

\subsubsection{Auxiliary conformal gauge theory of gravity}

Any torsion-free action quadratic in the conformal curvatures generically
reduces to a linear combination of the square of the conformal curvature
and the square of the dilatational curvature.
\[
S=\int\alpha\boldsymbol{\Omega}_{\;\;\;b}^{a}\land\,^{*}\boldsymbol{\Omega}_{\;\;\;a}^{b}+\beta\boldsymbol{\Theta}\land\,^{*}\boldsymbol{\Theta}
\]
Metric variation of $S$ gives the fourth-order, torsion-free Weyl
theory of gravity \cite{Weyl}. However, varying all 15 conformal
gauge fields we find additional field equations,

\begin{eqnarray*}
\alpha\mathbf{D}\,^{*}\boldsymbol{\Omega}_{\;\;\;b}^{a} & = & 0\\
\beta\mathbf{d}\,^{*}\boldsymbol{\Theta} & = & 0\\
2\alpha\mathbf{e}^{b}\wedge\,^{*}\boldsymbol{\Omega}_{\;\;\;b}^{a}-\beta\mathbf{e}^{a}\wedge\,^{*}\boldsymbol{\Theta} & = & 0\\
-\alpha\mathsf{f}_{ac}\left(\Omega^{abcd}+\Omega^{adcb}\right)+\frac{1}{2}\beta\left(\mathsf{f}_{\;\;\;c}^{b}\Theta^{cd}+\mathsf{f}_{\;\;\;c}^{d}\Theta^{cb}\right) & = & \alpha\Sigma^{bd}+\beta T^{bd}
\end{eqnarray*}
The sources on the right have the form of energy tensors for the Weyl
curvature and dilatational curvature. The first, $\Sigma_{ab}=-\left(\Omega_{\;\;\;a}^{cde}\Omega_{cdeb}-\frac{1}{4}\eta_{ab}\Omega^{cdef}\Omega_{cdef}\right)=0$,
vanishes by an identity first shown by Lanczos \cite{Lanczos} and
also found in \cite{Lovelock}. The second source term $T_{ab}=\Theta_{ac}\Theta_{b}^{\;\;\;c}-\frac{1}{4}\eta_{ab}\Theta_{cd}\Theta^{cd}=0$
vanishes because the dilatational curvature is zero, $\boldsymbol{\Theta}=0$.

Using the results in the preceding Subsection the field equations
reduce to
\begin{eqnarray}
\mathbf{D}\,^{*}\mathbf{C}_{\;\;\;b}^{a} & = & 0\label{Vanishing Div C}\\
2\alpha\mathbf{e}^{b}\wedge\,^{*}\mathbf{C}_{\;\;\;b}^{a} & = & 0\label{C trace-free}\\
-\alpha\mathcal{R}_{ac}\left(C^{abcd}+C^{adcb}\right) & = & 0\label{C R contraction symmetric}
\end{eqnarray}
The vanishing divergence of Eq.(\ref{Vanishing Div C}) is the vanishing
of the Cotton tensor, a necessary but not sufficient condition for
conformal flatness. The second equation, Eq.(\ref{C trace-free})
just expresses the tracelessness of the Weyl curvature. Together,
the final symmetry requirement (\ref{C R contraction symmetric})
and the vanishing Cotton tensor have been shown to imply the vanishing
of the Einstein tensor of an integrable Weyl geometry \cite{Wheeler 2014}.
This happens because when the full conformal connection is varied
(instead of just the metric) these two additional field equations
are integrability conditions that reduce the solutions to the familiar
second order form.

This brings us full circle, reproducing the conclusions of \cite{EPS original,EPS,Matveev1,Matveev2}
without their lightlike limit, but requiring a suitable theory of
gravity.

\section{A brief summary}

We have extended studies based on the Ehlers, Pirani, Schild program,
without the assumption of a symmetric, metric compatible connection.
The increased generality results in a 1-parameter class of projective
transformations, Eq.(\ref{Projective transformation})
\[
\tilde{\Sigma}_{\left(a\right)\;\mu\nu}^{\;\alpha}=\Sigma_{\;\;\;\mu\nu}^{\alpha}+a\delta_{\mu}^{\alpha}\xi_{\nu}+\left(1-a\right)\delta_{\nu}^{\alpha}\xi_{\mu},\quad\forall a\in\mathbb{R}
\]
We examined the effect of every member of this class on the curvature
and torsion. Introducing an arbitrary, nondegenerate, symmetric tensor-valued
functional as a metric allowed us to access the remaining degrees
of freedom of a general connection. Allowing nonzero nonmetricity
removes any obstruction from regarding this functional as the metric.
From the changes induced by projective transformations on inner products,
torsion, nonmetricity, and curvature, we revised Weyl's expression
for the projectively invariant curvature to
\begin{eqnarray*}
\mathcal{R}_{\;\;\;\beta\mu\nu}^{\alpha} & = & R_{\;\;\;\beta\mu\nu}^{\alpha}+\frac{1}{n-1}\left(\delta_{\nu}^{\alpha}R_{\beta\mu}-\delta_{\mu}^{\alpha}R_{\beta\nu}\right)-\left(1-a\right)\mathcal{W}_{\alpha}\mathcal{T}_{\;\;\;\mu\nu}^{\alpha}
\end{eqnarray*}
and found additional invariant expressions
\begin{eqnarray*}
\mathcal{T}_{\;\;\;\beta\mu}^{\alpha} & = & T_{\;\;\;\beta\mu}^{\alpha}+\frac{1}{n-1}\left(\delta_{\mu}^{\alpha}T_{\;\;\;\nu\beta}^{\nu}-\delta_{\beta}^{\alpha}T_{\;\;\;\nu\mu}^{\nu}\right)\\
\mathcal{Q}_{\;\;\;\beta\mu}^{\alpha} & = & Q_{\;\;\;\beta\mu}^{\alpha}-\left(2a\delta_{\beta}^{\alpha}\mathcal{W}_{\mu}+\left(1-a\right)\left(\delta_{\mu}^{\alpha}\mathcal{W}_{\beta}+g_{\beta\mu}\mathcal{W}^{\alpha}\right)\right)\\
\Phi_{\mu\nu} & = & D_{\nu}\mathcal{W}_{\mu}-D_{\mu}\mathcal{W}_{\nu}+\mathcal{W}_{\sigma}T_{\;\;\;\mu\nu}^{\sigma}
\end{eqnarray*}
where $\Phi_{\mu\nu}$ generalizes the dilatational curvature to an
affine geometry.

From the effect of regional projections on the resulting nonmetricity
and metric, we showed that it is also possible to build an equivalent
invariant geometry using a conformal class of metrics. We wrote conformally
invariant forms for the curvature, torsion, and nonmetricity. 

The projective and conformal connections depend on the \emph{same}
gauge vector. While the projective and conformal transformations have
different effects on that gauge vector, either type of transformation
can remove the vector and simultaneously reduce \emph{both} the projective
and conformal geometries to the original form. By comparing the conditions
for projective Ricci flatness and conformal Ricci flatness, we showed
that each implies the other.

We showed that both the projective and the conformal options require
the introduction of the same monotonic, twice differentiable function
on a region of spacetime and found that existence of such a function
requires a region foliated by order isomorphic, totally ordered, $C^{2}$
timelike curves in a geometry with only a single timelike dimension,
i.e. Lorentzian.

Having found that measurement of both timelike and null geodesics
leads to a conformal spacetime, we corrected the original Ehlers,
Pirani, and Schild restriction to a Weyl connection to include a full
conformal connection. This is the most we can determine from scalar
and spinning particles moving on extremal curves. To restrict the
geometry further requires a theory of gravity based on at least an
$SO\left(4,2\right)$ connection. We wrote the structure equations
and showed that they already imply the auxiliary character of the
gauge field of special conformal transformations. Finally, assuming
the simplest form of the action using this connection, we summarized
how reductions of the system lead us back to an integrable Weyl geometry.

\end{document}